\documentclass[runningheads]{llncs}
\usepackage{graphicx}
\usepackage{lscape}
\usepackage[table,xcdraw]{xcolor}
\usepackage{hyperref}
\usepackage{subcaption}
\usepackage{xcolor}
\usepackage{soul}
\usepackage{rotating}
\usepackage{pifont}
\usepackage[percent]{overpic}
\usepackage{amsfonts} 
\usepackage{mathtools} 
\usepackage{makecell}
\makeatletter
\newcommand{\printfnsymbol}[1]{%
  \textsuperscript{\@fnsymbol{#1}}%
}
\makeatother

\begin{document}
\title{Overview of the HECKTOR Challenge at MICCAI 2021: Automatic Head and Neck Tumor Segmentation and Outcome Prediction in PET/CT Images}
\titlerunning{Overview of the HECKTOR 2021 Challenge}
%
\author{Vincent Andrearczyk\thanks{equal contribution}\inst{1} \and
Valentin Oreiller\printfnsymbol{1}\inst{1,2} \and
Sarah Boughdad\inst{2} \and
Catherine Cheze Le Rest\inst{3,6} \and
Hesham Elhalawani\inst{4} \and
Mario Jreige\inst{2} \and
John O. Prior\inst{2} \and
Martin Valli\`eres\inst{5} \and
Dimitris Visvikis\inst{6}\and
Mathieu Hatt\thanks{equal contribution}\inst{6}\and
Adrien Depeursinge\printfnsymbol{2}\inst{1,2}}
\authorrunning{V. Andrearczyk et al.}
%

\institute{Institute of Information Systems, School of Management, HES-SO Valais-Wallis University of Applied Sciences and Arts Western Switzerland, Sierre, Switzerland
\and Centre Hospitalier Universitaire Vaudois (CHUV), Lausanne, Switzerland
\and Centre Hospitalier Universitaire de Poitiers (CHUP), Poitiers, France
\and Cleveland Clinic Foundation, Department of Radiation Oncology, Cleveland, OH, USA\\
\and Department of Computer Science, University of Sherbrooke, Sherbrooke, Québec, Canada
\and LaTIM, INSERM, UMR 1101, Univ Brest, Brest, France
}
\let\oldmaketitle\maketitle
\renewcommand{\maketitle}{\oldmaketitle\setcounter{footnote}{0}}

\maketitle              
\begin{abstract}
This paper presents an overview of the second edition of the HEad and neCK TumOR (HECKTOR) challenge, organized as a satellite event of the 24th International Conference on Medical Image Computing and Computer Assisted Intervention (MICCAI) 2021. The challenge is composed of three tasks related to the automatic analysis of PET/CT images for patients with Head and Neck cancer (H\&N), focusing on the oropharynx region. \textit{Task 1} is the automatic segmentation of H\&N primary Gross Tumor Volume (GTVt) in FDG-PET/CT images. \textit{Task 2} is the automatic prediction of Progression Free Survival (PFS) from the same FDG-PET/CT. Finally, \textit{Task 3} is the same as Task 2 with ground truth GTVt annotations provided to the participants.
The data were collected from six centers for a total of 325 images, split into 224 training and 101 testing cases. The interest in the challenge was highlighted by the important participation with 103 registered teams and 448 result submissions. The best methods obtained a Dice Similarity Coefficient (DSC) of 0.7591 in the first task, and a Concordance index (C-index) of 0.7196 and 0.6978 in Tasks 2 and 3, respectively.
In all tasks, simplicity of the approach was found to be key to ensure generalization performance.  
The comparison of the PFS prediction performance in Tasks 2 and 3 suggests that providing the GTVt contour was not crucial to achieve best results, which indicates that fully automatic methods can be used.
This potentially obviates the need for GTVt contouring, opening avenues for reproducible and large scale radiomics studies including thousands potential subjects.

\keywords{Automatic Segmentation\and Challenge\and Medical Imaging \and Head and Neck Cancer \and Segmentation \and Radiomics \and Deep Learning}
\end{abstract}
\section{Introduction: Research Context}

The prediction of disease characteristics and outcomes using quantitative image biomarkers from medical images (i.e. radiomics) has shown tremendous potential to optimize and personalize patient care, particularly in the context of Head and Neck (H\&N) tumors \cite{vallieres2017radiomics}.
FluoroDeoxyGlucose (FDG)-Positron Emission Tomography (PET) and Computed Tomography (CT) imaging are the modalities of choice for the initial staging and follow-up of H\&N cancer, as well as for radiotherapy planning purposes.
Yet, both gross tumor volume (GTV) delineations in radiotherapy planning and radiomics analyses aiming at predicting outcome rely on an expensive and error-prone manual or semi-automatic annotation process of Volumes of Interest (VOI) in three dimensions. The fully automatic segmentation of H\&N tumors from FDG-PET/CT images could therefore enable faster and more reproducible GTV definition as well as the validation of radiomics models on very large cohorts. 
Besides, fully automatic segmentation algorithms could also facilitate the application of validated models to patients' images in routine clinical workflows.
By focusing on metabolic and morphological tissue properties respectively, PET and CT images provide complementary and synergistic information for cancerous lesion segmentation and patient outcome prediction. 
The HEad and neCK TumOR segmentation and outcome prediction from PET/CT images (HECKTOR)\footnote{\url{https://www.aicrowd.com/challenges/miccai-2021-hecktor}, as of October 2021.} challenge aimed at identifying the best methods to leverage the rich bi-modal information in the context of H\&N primary tumor segmentation and outcome prediction. The analysis of the results provides precious information on the adequacy of the image analysis methods for the different tasks and the feasibility of large-scale and reproducible radiomics studies. 

The potential of PET information for automatically segmenting tumors has been long exploited in the literature. For an in-depth review of automatic segmentation of PET images in the pre-deep learning era, see \cite{foster2014review,hatt2017classification} covering methods such as fixed or adaptive thresholding, active contours, statistical learning, and mixture models. The need for a standardized evaluation of PET automatic segmentation methods and a comparison study between all the current algorithms was highlighted in \cite{hatt2017classification}.
The first challenge on tumor segmentation in PET images was proposed at MICCAI 2016\footnote{\url{https://portal.fli-iam.irisa.fr/petseg-challenge/overview\#_ftn1}, as of October 2020.} by Hatt et al \cite{Hatt2018}, implementing evaluation recommendations published previously by the AAPM (American Association of Physicists in Medicine) Task group 211 \cite{hatt2017classification}. 
Multi-modal analyses of PET and CT images have also recently been proposed for different tasks, including lung cancer segmentation in \cite{kumar2019co,li2019deep,zhao2018tumor,zhong20183d} and bone lesion detection in \cite{xu2018automated}.
In \cite{andrearczyk2020automatic}, we developed a baseline Convolutional Neural Network (CNN) approach based on a leave-one-center-out cross-validation on the training data of the HECKTOR challenge. Promising results were obtained with limitations that motivated additional data curation, data cleaning and the creation of the first HECKTOR challenge in 2020~\cite{andrearczyk2020overview,oreiller2021head}. 
This first edition compared segmentation architectures as well as the complementarity of the two modalities for the segmentation of GTVt in H\&N. 

In this second edition of the challenge, we propose a larger dataset, including a new center to better evaluate the generalization of the algorithms, as well as new tasks of prediction of Progression-Free Survival (PFS). 
Preliminary studies of automatic PFS prediction were performed with standard radiomics~\cite{fontaine2021fully} and deep learning models~\cite{andrearczyk2021multi} prior to challenge design.
The proposed dataset comprises data from six centers. Five centers are used for the training data and two for testing (data from the sixth center are split between training and testing sets). 
The task is challenging due to, among others, the variation in image acquisition and quality across centers (the test set contains data from a domain not represented in the training set) and the presence of lymph nodes with high metabolic responses in the PET images. 

The critical consequences of the lack of quality control in challenge designs were shown in~\cite{maier2018rankings} including reproducibility and interpretation of the results often hampered by the lack of provided relevant information and the use of non-robust ranking of algorithms. Solutions were proposed in the form of the Biomedical Image Analysis challengeS (BIAS)~\cite{maier2020bias} guidelines for reporting the results.
This paper presents an overview of the challenge following these guidelines.

Individual participants' papers reporting their methods and results were submitted to the challenge organizers. Reviews were organized by the organizers and the papers of the participants are published in the LNCS challenges proceedings~\cite{wang2022ccut,an2022coarse,qayyum2022automatic,starke2022hybrid,wang2022head,cho2022multimodal,ren2022pet,xie2022head,ghimire2022head,lang2022deep,meng2022multi,liu2022unet,saeed2022ensemble,asenjo2022pet,yuan2022automatic,de2022skip,bourigault2022pet,naser2022head,naser2022progression,wahid2022combining,ma2022self,paeenafrakati2022advanced,fatan2022fusion,murugesan2022head,yousefirizi2022segmentation,lu2022priori,lee2022dual,muller2022deep,ren2022comparing}. When participating in multiple tasks, participants could submit one or multiple papers.

The manuscript is organized as follows. 
The challenge dataset is described in Section~\ref{sec:dataset}.
The tasks descriptions, including challenge design, algorithms summaries and results, are split into two sections. The segmentation task (Task 1) is presented in Section~\ref{sec:task1}, and the outcome prediction tasks (Tasks 2 and 3) are described in Section~\ref{sec:task23} .
Section~\ref{sec:discussion} discusses the results and findings and Section~\ref{sec:conclusion} concludes the paper.

\section{Dataset}\label{sec:dataset}

\subsection{Mission of the Challenge}

\paragraph{Biomedical application}\mbox{}
\\The participating algorithms target the following fields of application: diagnosis, prognosis and research.
The participating teams' algorithms were designed for either or both image segmentation (i.e., classifying voxels as either primary tumor, GTVt, or background) and PFS prediction (i.e., ranking patients according to predicted risk of progression).

\paragraph{Cohorts}\mbox{}
\\As suggested in~\cite{maier2020bias}, we refer to the patients from whom the image data were acquired as the challenge cohort.
The target cohort\footnote{The target cohort refers to the subjects from whom the data would be acquired in the final biomedical application. It is mentioned for additional information as suggested in BIAS, although all data provided for the challenge are part of the challenge cohort.} comprises patients received for initial staging of H\&N cancer. 
The clinical goals are two-fold; the automatically segmented regions can be used as a basis for (i) treatment planning in radiotherapy, (ii) further radiomics studies to predict clinical outcomes such as overall patient survival, disease-free survival, response to therapy or tumor aggressiveness. Note that the PFS outcome prediction task does not necessarily have to rely on the output of the segmentation task.
In the former case (i), the regions will need to be further refined or extended for optimal dose delivery and control.
The challenge cohort\footnote{The challenge cohort refers to the subjects from whom the challenge data were acquired.} includes patients with histologically proven H\&N cancer who underwent radiotherapy treatment planning. The data were acquired from six centers (four for the training, one for the testing, and one for both) with variations in the scanner manufacturers and acquisition protocols. 
The data contain PET and CT imaging modalities as well as clinical information including age, sex, acquisition center, TNM staging, HPV status and alcohol. A detailed description of the annotations is provided in Section~\ref{subsec:dataset}.

\paragraph{Target entity}\mbox{}
\\The data origin, i.e. the region from which the image data were acquired, varied from the head region only to the whole body. While we provided the data as acquired, we limited the analysis to the oropharynx region and provided a semi-automatically detected bounding-box locating the oropharynx region~\cite{andrearczyk2020oropharynx}, as illustrated in Fig.~\ref{fig:examples_cases}. The participants could use the entire images if wanted but the predictions were evaluated only within these bounding-boxes. 

\subsection{Challenge Dataset}\label{subsec:dataset}
\paragraph{Data source}\mbox{}
\\The data were acquired from six centers as detailed in Table~\ref{tab:listCenters}. It consists of PET/CT images of patients with H\&N cancer located in the oropharynx region.
The devices and imaging protocols used to acquire the data are described in Table~\ref{tab:listDevices}.
Additional information about the image acquisition is provided in Appendix~\ref{app:acquisition}.

\begin{table}
\caption{List of the hospital centers in Canada (CA), Switzerland (CH) and France (FR) and number of cases, with a total of 224 training and 101 test cases.}\label{tab:listCenters}
\centering
\begin{tabular}{l|c|c}
Center &  Split & \# cases\\
\hline
HGJ: Hôpital Général Juif, Montréal, CA & Train  & 55 \\
CHUS: Centre Hospitalier Universitaire de Sherbooke, Sherbrooke, CA & Train & 72 \\
HMR: Hôpital Maisonneuve-Rosemont, Montréal, CA & Train & 18 \\
CHUM: Centre Hospitalier de l’Université de Montréal, Montréal, CA & Train & 56 \\
CHUP: Centre Hospitalier Universitaire Poitiers, FR & Train & 23 \\
\hline
Total & Train & 224 \\
\hline
CHUV: Centre Hospitalier Universitaire Vaudois, CH & Test & 53 \\
CHUP: Centre Hospitalier Universitaire Poitiers, FR & Test & 48 \\
\hline
Total & Test & 101 \\
\end{tabular}
\end{table}

\begin{table}
\caption{List of scanners used in the various centers.}\label{tab:listDevices}
\centering
\begin{tabular}{c|c}
Center &  Device\\
\hline
HGJ & hybrid PET/CT scanner (Discovery ST, GE Healthcare) \\
CHUS & hybrid PET/CT scanner (Gemini GXL 16, Philips) \\
HMR & hybrid PET/CT scanner (Discovery STE, GE Healthcare) \\
CHUM & hybrid PET/CT scanner (Discovery STE, GE Healthcare) \\
CHUV & hybrid PET/CT scanner (Discovery D690 ToF, GE Healthcare) \\
CHUP & hybrid PET/CT scanner (Biograph mCT 40 ToF, Siemens) \\
\end{tabular}
\end{table}

\begin{figure}
\centering
\begin{subfigure}{.45\textwidth}
  \centering
  \includegraphics[width=0.8\linewidth]{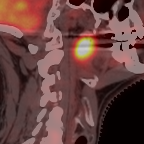}
  \caption{CHUM}
\end{subfigure}%
\begin{subfigure}{.45\textwidth}
  \centering
  \includegraphics[width=0.8\linewidth]{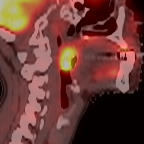}
  \caption{CHUS}
\end{subfigure}
\begin{subfigure}{.45\textwidth}
  \centering
  \includegraphics[width=0.8\linewidth]{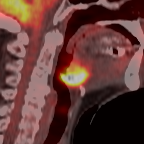}
  \caption{HGJ}
\end{subfigure}
\begin{subfigure}{.45\textwidth}
  \centering
  \includegraphics[width=0.8\linewidth]{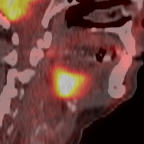}
  \caption{HMR}
\end{subfigure}
\begin{subfigure}{.45\textwidth}
  \centering
  \includegraphics[width=0.8\linewidth]{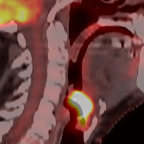}
  \caption{CHUV}
\end{subfigure}
\begin{subfigure}{.45\textwidth}
  \centering
  \includegraphics[width=0.8\linewidth]{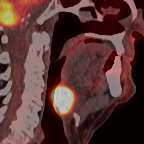}
  \caption{CHUP}
\end{subfigure}
\caption{Case examples of 2D sagittal slices of fused PET/CT images from each of the six centers. The CT (grayscale) window in Hounsfield unit is $[-140, 260]$ and the PET window in SUV is $[0, 12]$, represented in a "hot" colormap.}
\label{fig:examples_cases}
\end{figure}

\paragraph{Training and test case characteristics}\mbox{}
\\The training data comprise 224 cases from five centers (HGJ, HMR\footnote{For simplicity, these centers were renamed CHGJ and CHMR during the challenge.}, CHUM, CHUS and CHUP). Originally, the dataset in~\cite{vallieres2017radiomics} contained 298 cases, among which we selected the cases with oropharynx cancer.  
The test data contain 101 cases from a fifth center CHUV (n=53) and CHUP (n=48). Examples of PET/CT images of each center are shown in Fig.~\ref{fig:examples_cases}.
Each case includes a CT image, a PET image and a GTVt mask (for the training cases) in the Neuroimaging Informatics Technology Initiative (NIfTI) format, as well as patient information (e.g. age, sex) and center. 
A bounding-box of size $144\times 144\times 144$ mm$^3$ locating the oropharynx region was also provided. Details of the semi-automatic region detection can be found in~\cite{andrearczyk2020oropharynx}. 

Finally, to provide a fair comparison, participants who wanted to use additional external data for training were asked to also report results using only the HECKTOR data and discuss differences in the results. 
However, no participant used external data in this edition.

\paragraph{Annotation characteristics}\mbox{}\\
For the HGJ, CHUS, HMR, and CHUM centers, initial annotations, i.e. 3D contours of the GTVt, were made by expert radiation oncologists and were later re-annotated as described below.
Details of the initial annotations of the training set can be found in ~\cite{vallieres2017radiomics}. In particular, 40\% (80 cases) of the training radiotherapy contours were directly drawn on the CT of the PET/CT scan and thereafter used for treatment planning. 
The remaining 60\% of the training radiotherapy contours were drawn on a different CT scan dedicated to treatment planning and were then registered to the FDG-PET/CT scan reference frame using intensity-based free-form deformable registration with the software MIM (MIM software Inc., Cleveland, OH). The initial contours of the test set were all directly drawn on the CT of the PET/CT scan.

The original contours for the CHUV center were drawn by an expert radiation oncologist for radiomics purpose~\cite{castelli2019pet}. The expert contoured the tumors on the PET/CT scan.
The delineation from the CHUP center were obtained semi-automatically with a Fuzzy Locally Adaptive Bayesian (FLAB) segmentation~\cite{hatt2009fuzzy} and corrected by an expert radiation oncologist. These contours were obtained on the PET images only.

Given the heterogeneous nature of the original contours, a re-annotation of the VOIs was performed. During the first edition of HECKTOR (HGJ, CHUS, HMR, CHUM, and CHUV), the re-annotation was supervised by an expert who is both radiologist and nuclear medicine physician.
Two non-experts (organizers of the challenge) made an initial cleaning in order to facilitate the expert's work.
The expert either validated or edited the VOIs.
The Siemens Syngo.Via RT Image Suite was used to edit the contours in 3D with fused PET/CT images. Most of the contours were re-drawn completely, and the original segmentations were used to localize the tumor and discriminate between malignant versus benign high metabolic regions.

For the data added to the current HECKTOR edition (CHUP), the re-annotation was performed by three experts: one nuclear medicine physician, one radiation oncologist and one who is both radiologist and nuclear medicine physician. The 71 cases were divided between the three experts and each annotation was then checked by all three of them. This re-annotation 
was performed in a centralized fashion with the MIM software, and the verification of the contours was made possible by the MIM Cloud platform \footnote{\url{https://mim-cloud.appspot.com/} as of December 2021.}.
Guidelines for re-annotating the images were developed by our experts and are stated in the following.

Oropharyngeal lesions are contoured on PET/CT using information from PET and unenhanced CT acquisitions. The contouring includes the entire edges of the morphologic anomaly as depicted on unenhanced CT (mainly visualized as a mass effect) and the corresponding hypermetabolic volume, using PET acquisition, unenhanced CT and PET/CT fusion visualizations based on automatic co-registration.
The contouring excludes the hypermetablic activity projecting outside the physical limits of the lesion (for example in the lumen of the airway or on the bony structures with no morphologic evidence of local invasion).
The standardized nomenclature per AAPM TG-263 is ``GTVp".
For more specific situations, clinical nodal category was verified to ensure the exclusion of nearby FDG-avid and/or enlarged lymph nodes (e.g. submandibular, high level II, and retropharyngeal).
In case of tonsillar fossa or base of tongue fullness/enlargement without corresponding FDG avidity, the clinical datasheet was reviewed to exclude patients with pre-radiation tonsillectomy or extensive biopsy.

\paragraph{Data preprocessing methods}\mbox{}
\\No preprocessing was performed on the images to reflect the diversity of clinical data and to leave full flexibility to the participants.
However, we provided various pieces of code to load, crop and resample the data, as well as to evaluate the results on our GitHub repository\footnote{\url{github.com/voreille/hecktor}, as of December 2021.}.
This code was provided as a suggestion to help the participants and to maximize transparency, but the participants were free to use other methods.

\paragraph{Sources of errors}\mbox{}
\\In~\cite{oreiller2021head}, we reported an inter-observer (four observers) agreement of 0.6110 on a subset of the HECKTOR 2020 data containing 21 randomly drawn cases. 
Similar agreements were reported in the literature~\cite{gudi2017interobserver} with an average DSC agreement of three observers of 0.57 using only the CT images for annotation and 0.69 using both PET and CT. 
A source of error therefore originates from the degree of subjectivity in the annotation and correction of the expert.
Another source of error is the difference in the re-annotation between the centers used in HECKTOR 2020 and the one added in HECKTOR 2021. In HECKTOR 2020, the re-annotation was checked by only one expert while for HECKTOR 2021 three experts participated in the re-annotation. Moreover, the softwares used were different.

Another source of error comes from the lack of CT images with a contrast agent for a more accurate delineation of the primary tumor.

\paragraph{Institutional review boards}\mbox{}
\\Institutional Review Boards (IRB) of all participating institutions permitted the use of images and clinical data, either fully anonymized or coded, from all cases for research purposes only. Retrospective analyses were performed following the relevant guidelines and regulations as approved by the respective institutional ethical committees with protocol numbers: MM-JGH-CR15-50 (HGJ, CHUS, HMR, CHUM) and CER-VD 2018-01513 (CHUV). In the case of CHUP, ethical review and approval were waived because data were already collected for routine patient management before analysis, in which patients provided informed consent. No additional data was specifically collected for the present challenge.

\section{Task 1: Segmentation}\label{sec:task1}
\subsection{Methods: Reporting of Challenge Design}\label{subsec:task1-methods}

A summary of the information on the challenge organization is provided in Appendix~\ref{app:challenge_info}, following the BIAS recommendations.

\paragraph{Assessment aim} \mbox{}
\\The assessment aim for the segmentation task is the following;
evaluate the feasibility of fully automatic GTVt segmentation for H\&N cancers in the oropharyngeal region via the identification of the most accurate segmentation algorithm. The performance of the latter is identified by computing the Dice Similarity Coefficient (DSC) and Hausdorff Distance (HD) at 95\textsuperscript{th} percentile (HD95) between prediction and manual expert annotations. 

DSC measures volumetric overlap between segmentation results and annotations. It is a good measure of segmentation for imbalanced segmentation problems, i.e. the region to segment is small as compared to the image size. DSC is commonly used in the evaluation and ranking of segmentation algorithms and particularly tumor segmentation tasks~\cite{gudi2017interobserver,moe2019deep}.

In 3D, the HD computes the maximal distance between the surfaces of two segmentations. It provides an insight on how close the boundaries of the prediction and the ground truth are. The HD95 measure the 95\textsuperscript{th} quantile of the distribution of surface distances instead of the maximum. This metric is more robust towards outlier segmented pixels than the HD and thus is often used to evaluate automatic algorithms~\cite{liu2020automatic,kuijf2019standardized}.

\paragraph{Assessment Method}\mbox{}
Participants were given access to the test cases without the ground truth annotations and were asked to submit the results of their
algorithms on the test cases on the AIcrowd platform. 

Results were ranked using the DSC and HD95, both computed on images cropped using the provided bounding-boxes (see Section~\ref{subsec:dataset}) in the original CT resolution. If the submitted results were in a resolution different from the CT resolution, we applied nearest-neighbor interpolation before evaluation.

The two metrics are defined for set $A$ (ground truth volumes) and set $B$ (predicted volumes) as follow:
\begin{equation}\label{eq:dice_score}
    \mathrm{DSC}(A,B) = \frac{2|A \cap B|}{|A| + |B|},
\end{equation}
where $|\cdot|$ is the set cardinality and 
\begin{equation}
\mathrm{HD95}(A,B)  = P_{95}\left\{ \adjustlimits \sup_{a\in A} \inf_{b\in B} \mathrm{d}(a,b), \adjustlimits \sup_{b\in B} \inf_{a\in A} \mathrm{d}(a,b)\right\},
\end{equation}

where $\mathrm{d}(a,b)$ is the Euclidean distance between points $a$ and $b$, $\sup$ and $\inf$ are the supremum and infimum, respectively. $P_{95}$ is the 95\textsuperscript{th} percentile.

The ranking was computed from the average DSC and median HD95 across all cases. Since the HD95 is unbounded, i.e. it is infinity when there is no prediction, we choose the median instead of the mean for aggregation. The two metrics are ranked separately and the final rank is obtained by Borda counting.
This ranking method was used first to determine the best submission of each participating team (ranking the 1 to 5 submissions), then to obtain the final ranking (across all participants).
Each participating team had the opportunity to submit up to five (valid) runs.
The final ranking is reported in Section~\ref{sec:results} and discussed in Section~\ref{sec:discussion}. 

Missing values (i.e. missing predictions on one or multiple patients), did not occur in the submitted results but would have been treated as DSC of zero and a HD95 of infinity.
In the case of tied rank (which was very unlikely due to the computation of the results average of 53 DSCs), we considered precision as the second ranking metric.
The evaluation implementation can be found on our GitHub repository\footnote{\url{github.com/voreille/hecktor}, as of December 2021.} and was made available to the participants to maximize transparency.

\subsection{Results: Reporting of Segmentation Task Outcome}\label{sec:results}

\subsubsection{Participation}\label{sec:participation}
At of September 14 2021 (submission deadline), the number of registered teams was 44 for Task 1, 30 for Task 2 and 8 for Task 3. A team is made of at least one participant and not all participants that signed the End User Agreement (EUA) registered a team. Each team could submit up to five valid submissions. By the submission deadline, we had received 448 results submissions, including valid and invalid ones (i.e. not graded due to format errors). This participation was much higher than last year's challenge with 83 submissions and highlights the growing interest in the challenge. 

In this section, we present the algorithms and results of participants who submitted a paper~\cite{muller2022deep,lee2022dual,lu2022priori,yousefirizi2022segmentation,murugesan2022head,fatan2022fusion,paeenafrakati2022advanced,naser2022head,bourigault2022pet,de2022skip,yuan2022automatic,asenjo2022pet,liu2022unet,meng2022multi,lang2022deep,ghimire2022head,xie2022head,ren2022pet,cho2022multimodal,wang2022head,qayyum2022automatic,an2022coarse,wang2022ccut}. An exhaustive list of the results can be seen on the leaderboard\footnote{\url{https://www.aicrowd.com/challenges/miccai-2021-hecktor/leaderboards?challenge_leaderboard_extra_id=667&challenge_round_id=879}}.

\subsubsection{Segmentation: summary of participants' methods}\label{sec:algoSegmentation}

This section summarizes the approaches proposed by all teams for the automatic segmentation of the primary tumor (Task 1).
Table~\ref{tab:algoSegmentation} provides a synthetic comparison of the methodological choices and design. 
All methods are further detailed in dedicated paragraphs.
The paragraphs are ordered according to the official ranking, starting with the winners of Task 1.
\begin{landscape}
\begin{table}[!ht]
    \centering
    \begin{tabular}{lll|llll|lllll|llll|llll|lllll|}
        ~ & ~ & ~ & \multicolumn{4}{c}{Preprocess.} & \multicolumn{5}{c}{\makecell{Data \\ augmentation}} & \multicolumn{4}{c}{\makecell{Model \\ archit.}} & \multicolumn{4}{c}{Loss} & \multicolumn{5}{c}{\makecell{Training/\\evaluation}} \\ 
        Team & Dice & HD95 & \rotatebox[origin=l]{90}{iso-resampling} & \rotatebox[origin=l]{90}{CT clipping} & \rotatebox[origin=l]{90}{Min-max norm.} & \rotatebox[origin=l]{90}{Standardization} & \rotatebox[origin=l]{90}{Rotation} & \rotatebox[origin=l]{90}{Scaling} & \rotatebox[origin=l]{90}{Flipping} & \rotatebox[origin=l]{90}{Noise addition} & \rotatebox[origin=l]{90}{Other} & \rotatebox[origin=l]{90}{U-Net} & \rotatebox[origin=l]{90}{Attention} & \rotatebox[origin=l]{90}{Res. connection} & \rotatebox[origin=l]{90}{SE norm.~\cite{iantsen2021squeeze}} & \rotatebox[origin=l]{90}{Dice } & \rotatebox[origin=l]{90}{Cross-entropy} & \rotatebox[origin=l]{90}{Focal~\cite{lin2017focal}} & \rotatebox[origin=l]{90}{Else} & \rotatebox[origin=l]{90}{Optimizer} & \rotatebox[origin=l]{90}{nnU-Net~\cite{isensee2021nnu}} & \rotatebox[origin=l]{90}{LR decay} & \rotatebox[origin=l]{90}{Cross-validation} & \rotatebox[origin=l]{90}{Ensembling}\\ \hline
        Pengy~\cite{xie2022head} & 0.7785 & 3.0882 & \checkmark & \checkmark & ~ & \checkmark & \checkmark & \checkmark & \checkmark & \checkmark & \checkmark & \checkmark & ~ & ~ & \checkmark & \checkmark & ~ & \checkmark & ~ & SGD & \checkmark & \checkmark & \checkmark & 5 \\ \hline
        SJTU EIEE 2-426Lab\footnote{It is ranked second due to the HD95 slightly better than the third (HiLab), 3.088160269617 vs 3.088161777508, and the ranking strategy describd in \ref{subsec:task1-methods}.} \cite{an2022coarse} & 0.7733 & 3.0882 & \checkmark & \checkmark & \checkmark & ~ & \checkmark & ~ & \checkmark & ~ & ~ & \checkmark & ~ & \checkmark & \checkmark & \checkmark & \checkmark & \checkmark & \checkmark & Adam & \checkmark & \checkmark & \checkmark & 9 \\ \hline
        HiLab \cite{lu2022priori}& 0.7735 & 3.0882 & \checkmark & \checkmark & ~ & \checkmark & ~ & ~ & ~ & ~ & ~ & \checkmark & \checkmark & ~ & \checkmark & \checkmark & ~ & \checkmark & ~ & Adam & ~ & \checkmark & \checkmark & 14 \\ \hline
        BCIOQurit \cite{yousefirizi2021gan}& 0.7709 & 3.0882 & \checkmark & \checkmark & ~ & \checkmark & \checkmark & \checkmark  & \checkmark & ~ & ~ & \checkmark & ~ & \checkmark & ~ & ~ & ~ & \checkmark & \checkmark & Adam & \checkmark & \checkmark & \checkmark & 10 \\ \hline
        Aarhus Oslo \cite{ren2022pet}& 0.7790 & 3.1549 & \checkmark & ~ & ~ & ~ & \checkmark & \checkmark & ~ & ~ & ~ & \checkmark & ~ & ~ & ~ & \checkmark & \checkmark & ~ & ~ & Adam & \checkmark & \checkmark & \checkmark & 3 \\ \hline
        Fuller MDA \cite{naser2022head}& 0.7702 & 3.1432 & \checkmark & \checkmark & \checkmark & \checkmark & \checkmark & \checkmark & \checkmark & ~ & \checkmark & \checkmark & ~ & \checkmark & ~ & \checkmark & ~ & ~ & ~ & Adam & ~ & \checkmark & \checkmark & 10 \\ \hline
        UMCG \cite{de2022skip}& 0.7621 & 3.1432 & \checkmark & ~ & ~ & \checkmark & ~ & ~ & ~ & ~ & ~ & \checkmark & \checkmark & ~ & ~ & ~ & ~ & \checkmark & \checkmark & Adam & ~ & \checkmark & \checkmark & 5 \\ \hline
        Siat \cite{wang2022head}& 0.7681 & 3.1549 & \checkmark & \checkmark & ~ & \checkmark & \checkmark & ~ & \checkmark & ~ & ~ & \checkmark & \checkmark & \checkmark & ~ & \checkmark & ~ & \checkmark & ~ & na & ~ & na & \checkmark & 5 \\ \hline
        Heck Uihak \cite{cho2022multimodal}& 0.7656 & 3.1549 & \checkmark & \checkmark & \checkmark & \checkmark & \checkmark & ~ & \checkmark & ~ & ~ & \checkmark & \checkmark & ~ & \checkmark & \checkmark & ~ & \checkmark & ~ & Adam & ~ & \checkmark & \checkmark & 5 \\ \hline
        BMIT USYD \cite{meng2022multi}& 0.7453 & 3.1549 & \checkmark & \checkmark & \checkmark & \checkmark & \checkmark & ~ & \checkmark & ~ & ~ & \checkmark & ~ & ~ & ~ & \checkmark & ~ & ~ & ~ & Adam & ~ & \checkmark & \checkmark & 10 \\ \hline
        DeepX \cite{yuan2022automatic}& 0.7602 & 3.2700 & \checkmark & \checkmark & ~ & \checkmark & ~ & ~ & \checkmark & ~ & \checkmark & \checkmark & \checkmark & \checkmark & \checkmark & \checkmark & ~ & ~ & ~ & Adam & ~ & \checkmark & \checkmark & 15 \\ \hline
        Emmanuelle Bourigault \cite{bourigault2022pet}& 0.7595 & 3.2700 & \checkmark & \checkmark & ~ & \checkmark & \checkmark & ~ & \checkmark & ~ & ~ & \checkmark & \checkmark & \checkmark & \checkmark & \checkmark & ~ & \checkmark & ~ & Adam & ~ & \checkmark & \checkmark & 5 \\ \hline
        C235 \cite{liu2022unet}& 0.7565 & 3.2700 & \checkmark & \checkmark & \checkmark & \checkmark & ~ & ~ & ~ & ~ & ~ & \checkmark & \checkmark & \checkmark & ~ & \checkmark & ~ & \checkmark & ~ & Adam & ~ & \checkmark & ~ & 5 \\ \hline
        Abdul Qayyum\cite{qayyum2022automatic}& 0.7487 & 3.2700 & ~ & ~ & \checkmark & ~ & ~ & ~ & \checkmark & ~ & \checkmark & \checkmark & ~ & \checkmark & ~ & \checkmark & \checkmark & ~ & ~ & Adam & ~ & ~ & \checkmark & ~ \\ \hline
        RedNeucon \cite{asenjo2022pet}& 0.7400 & 3.2700 & na & na & na & na & \checkmark & \checkmark & \checkmark & ~ & ~ & \checkmark & \checkmark & ~ & ~ & \checkmark & \checkmark & ~ & \checkmark & Adam & ~ & \checkmark & ~ & 25 \\ \hline
        DMLang \cite{lang2022deep}& 0.7046 & 4.0265 & \checkmark & \checkmark & \checkmark & ~ & \checkmark & ~ & \checkmark & \checkmark & \checkmark & \checkmark & ~ & ~ & ~ & \checkmark & ~ & ~ & ~ & Adam & ~ & ~ & ~ & ~ \\ \hline
        Xuefeng \cite{ghimire2022head}& 0.6851 & 4.1932 & \checkmark & \checkmark & ~ & \checkmark & \checkmark & \checkmark & \checkmark & \checkmark & ~ & \checkmark & ~ & ~ & ~ & \checkmark & \checkmark & ~ & ~ & SGD & ~ & \checkmark & \checkmark & ~ \\ \hline
        Qurit Tecvico \cite{paeenafrakati2022advanced}& 0.6771 & 5.4208 & ~ & ~ & \checkmark & ~ & ~ & ~ & ~ & ~ & ~ & \checkmark & ~ & ~ & ~ & \checkmark & ~ & ~ & ~ & Adam & ~ & ~ & ~ & ~ \\ \hline
        Vokyj \cite{muller2022deep}& 0.6331 & 6.1267 & \checkmark & \checkmark & ~ & \checkmark & \checkmark & ~ & ~ & ~ & ~ & ~ & ~ & ~ & ~ & \checkmark & ~ & \checkmark & ~ & Adam & ~ & \checkmark & ~ & ~ \\ \hline
        TECVICO Corp Family \cite{fatan2022fusion}& 0.6357 & 6.3718 & na & na & na & na & ~ & ~ & ~ & ~ & ~ & \checkmark & ~ & \checkmark & ~ & \checkmark & ~ & ~ & ~ & Adam & ~ & ~ & ~ & 2 \\ \hline\hline
        
        BAMF health \cite{murugesan2022head}& 0.7795 & 3.0571 & \checkmark & \checkmark & ~ & \checkmark & \checkmark & \checkmark & \checkmark & \checkmark & \checkmark & \checkmark & ~ & \checkmark & ~ & \checkmark & \checkmark & ~ & ~ & SGD & \checkmark & ~ & \checkmark & 10 \\ \hline
        Wangjiao \cite{wang2022ccut}& 0.7628 & 3.2700 & \checkmark & \checkmark & ~ & \checkmark & ~ & ~ & ~ & ~ & ~ & \checkmark & \checkmark & \checkmark & \checkmark & \checkmark & ~ & \checkmark & ~ & Adam & ~ & \checkmark & \checkmark & 6 \\ \hline
    \end{tabular}
    \caption{Synthetic comparison of segmentation methods and results. More details are available in Section~\ref{sec:algoSegmentation}. The number of used models is reported in the last column when ensembling was used. ``na" stands for ``not available".}\label{tab:algoSegmentation}
\end{table}
\end{landscape}

In~\cite{xie2022head}, Xie and Peng (team ``Pengy'') used a well-tuned patch-based 3D nnU-Net~\cite{isensee2021nnu} with standard pre-processing and training scheme, where the learning rate is adjusted dynamically using polyLR~\cite{ChenPK0Y16}.
The Squeeze and Excitation (SE) normalization~\cite{iantsen2021squeeze} was also one of the main ingredient of their approach.
The approach is straighforward yet efficient as they ranked first for Task 1. 
Five models are trained in a five-fold cross-validation with random data augmentation including rotation, scaling,
mirroring, Gaussian noise and Gamma correction. The five test predictions are ensembled via probability averaging for the final results.

In~\cite{an2022coarse}, An et al. (team ``SJTU EIEE 2-426Lab'') proposed a framework which is based on the subsequent application of three different U-Nets. 
The first U-Net is used to coarsely segment the tumor and then select a bounding-box.
Then, the second network performs a finer segmentation on the smaller bounding box.
Finally, the last network takes as input the concatenation of PET, CT and the previous segmentation to refine the predictions.
They trained the three networks with different objectives. 
The first one was trained to optimize the recall rate, and the two subsequent ones were trained to optimize the Dice score. 
All objectives were implemented with the F-loss which includes a hyper-parameter
allowing to balance between recall and Dice. The final prediction was obtained through majority voting on three different predictions: an ensemble of five nnU-Nets \cite{isensee2021nnu} (trained on five different folds), an ensemble of three U-Nets with SE     normalization~\cite{iantsen2021squeeze}, and the predictions made by the proposed model.

In~\cite{lu2022priori}, Lu et al. (team ``HiLab'') employed an ensemble of various 3D U-Nets, including the eight models used in~\cite{iantsen2021squeeze}, winner of HECKTOR 2020, five models trained with leave-one-center-out, and one model combining a priori and a posteriori attention. In this last model, the normalized PET image was used as a priori attention map for segmentation on the CT image. Mix-up was also used, mixing PET and CT in the training set to construct a new domain to account for the domain shift in the test set. All 14 predictions were averaged and thresholded to 0.5 for the final ensembled prediction.

In~\cite{yousefirizi2022segmentation}, Yousefirizi et al. (team ``BCIOqurit'') used a 3D nnU-Net with SE normalization~\cite{iantsen2021squeeze} trained on a leave-one-center-out with a combination of a "unified" focal and Mumford-Shah~\cite{kim2019mumford} losses taking the advantage of distribution, region, and boundary-based loss functions.

In~\cite{ren2022pet}, Ren et al. (team ``Aarhus Oslo'') proposed a 3D nnU-Net with various PET normalization methods, namely PET-clip and PET-sin. 
The former clips the Standardized Uptake Values (SUV) range in [0,5] and the latter transforms monotonic spatial SUV increase into onion rings via a sine transform of SUV.
Loss functions were also combined and compared (Dice, Cross-Entropy, Focal and TopK). 
No strong global trend was observed on the influence of the normalization or loss.

In~\cite{naser2022head}, Naser et al. (team ``Fuller MDA'') used an ensemble of 3D residual U-Nets trained on a 10-fold CV resulting in 10 different models. The ensemble was performed either by STAPLE or majority voting on the binarized predictions. Models with different numbers of channels were also compared.
The best combination was the one with fewer feature maps and ensembled with majority voting.

In~\cite{de2022skip}, De Biase et al. (team ``UMCG'') compared two methods: (i) Co-learning Multi-Modal PET/CT adapted from~\cite{Xue2021}, which takes as input PET and CT as two separate images, outputs two masks that are averaged and (ii) Skip-scSE Multi-Scale Attention, which concatenates PET and CT in the channel dimension.
The Skip-scSE models clearly outperformed the other. Ensembling (i) and (ii) provided worse results.

In~\cite{wang2022head}, Wang et al. (team ``Siat'') used an ensemble of 3D U-Nets with multi-channel attention mechanisms. For each channel in the input data, this attention module outputs a weighted combination of filter outputs from three receptive fields over the input. A comparison with a standard 3D Vnet without attention showed the benefit of the latter.

In \cite{cho2022multimodal}, Cho et al. team ``Heck Uihak'') used a backbone 3D U-Net that takes as input PET/CT images and outputs the predictions. This backbone U-Net is coupled with an attention module. The attention module was designed around a U-Net architecture and takes as input the PET images and produces attention maps. These attention maps are then multiplied with the skip connections of the backbone U-Net. The whole pipeline was trained with a sum of a Dice loss and a focal loss.

In~\cite{meng2022multi}, Meng et al. (team ``BMIT USYD'') used multi-task learning scheme to address Tasks 1 and 2. A modified 3D U-Net was used for segmentation. Its output is a voxel-wise tumor probability that is fed together with PET/CT to a 3D denseNet. Ensembling was used to produce the final output.

In~\cite{yuan2022automatic}, Yuan et al. (team ``DeepX'') proposed a 3D U-Net with scale attention which is referred to as Scale Attention Network (SA-Net). 
The skip connections were replaced by an attention block and the concatenation was replaced by a summation. The attention block takes as input the skip connections at all the scales and output an attention map which is added to the feature maps of the decoder. The attention blocks include a SE block. The encoder and decoder include ResNet-like blocks containing a SE block.
An ensemble of 15 models was used for the final prediction (5 from the 5-fold CV
with input size $144\times 144\times 144$ at 1mm$^3$, 5 from the 5-fold CV
with input size $128\times 128 \times 128$ at $1.25\times  1.25\times 1.25$mm$^3$, and 5 from a leave-one-center-out CV with input size $144\times 144\times 144$ at 1mm$^3$).

In~\cite{bourigault2022pet}, Bourigault et al.  (team ``Emmanuelle Bourigault'') proposed a full scale 3D U-Net architecture with attention, residual connections and SE norm. Conditional random fields was applied as post-processing.

In~\cite{liu2022unet}, the authors (team ``C235'') proposed a model based on 3D U-Net supplemented with a simple attention module referred to as SimAM. Different from channel-wise and spatial-wise attention mechanisms, SimAM generates the corresponding weight for each pixel in each channel and spatial position.
They compared their model to last year's winning algorithm based on SE Norm and report a small but consistent increase in segmentation performance when using the proposed SimAM attention module, which also resulted in models with about 20 times less parameters.

In~\cite{qayyum2022automatic}, Qayyum et al. (team ``Abdul Qayyum'') proposed to use a 3D U-Net with 3D inception as well as squeeze and excitation modules with residual connections. 
They extended the 2D inception module into 3D with extra 3D depth-wise layers for semantic segmentation.
The comparison with and without the inception module showed a systematic improvement associated with the latter.

In~\cite{asenjo2022pet}, Asenjo et al. (team ``RedNeucon'') ensembled a total of 25 models: a combination of 2D (trained on axial, sagittal and coronal planes) and 3D U-Nets, all trained on cross-validation and on the full dataset.

In~\cite{lang2022deep}, Lang et al. (team ``DMLang'') used a network based on a 3D U-Net. The main modification is that the skip connections were linked directly after the downsampling.
They also optimized the kernel size and the strides of the convolutions.

In \cite{ghimire2022head}, Ghimire et al. (team ``Xuefeng'') developed a patch-based 3D U-Net with overlapping sliding window at test time.
Deep supervision technique was applied to the network, where the computation of loss occurs at each decoding block. Various patch sizes, modality combination and convolution types were compared.
Results suggest that larger patch size, bi-modal inputs, and conventional convolution (i.e. not dilated) was better. 

In~\cite{paeenafrakati2022advanced} Paeenafrakati et al. (team ``Qurit Tecvico'') proposed to use 3D U-Net or 3D U-NeTr (U-Net with transformers) to segment the GTVt. The network's input consists of a one-channel image. This image was obtained by image-level fusion techniques to combine information of both PET and CT images. They assessed ten different image fusion methods.
To select the best combination of architecture and fusion method, they used a validation set of 23 images. The best combination was a U-Net architecture with the Laplacian pyramid method for fusion. This model obtained a DSC of, respectively, 0.81 and 0.68 on the validation and test set.

In~\cite{muller2022deep}, Muller et al. (team ``Vokyj'') proposed a model trained on supervoxels (obtained with Simple Linear Iterative Clustering, SLIC), motivated by the efficiency of the latter. The model is composed of an MLP encoder and graph CNN decoder. The models were trained on extracted patches of size 72x72x72.

In~\cite{fatan2022fusion}, Fatan et al. (team ``TECVICO Corp Family'') employed a 3D U-Net with autoencoder regularization~\cite{myronenko20183d} trained on various fusions of PET and CT images. The best results were obtained with a Laplacian pyramid-sparse representation mixture. 


Lee et al.~\cite{lee2022dual} (team ``Neurophet'') used a dual path encoder (PET, CT) whose paths are coupled by a shared-weight cross-information module in each layer of the encoding path of the 3D U-Net architecture.
The cross-attention module performs global average pooling over the feature channels resulting from convolutional blocks in both paths and feeds the resulting pooled features into a weight-shared fully connected layer. Its output, two (transformed) feature vectors are added elementwise and activated using a sigmoid function. The final output of each layer in the encoding part is obtained by multiplication of the features in each of the two paths with these cross-attention weights.
The study used the generalized dice loss as training metric. Five separate models were built, using data from four centers for training and data from the 5th center for evaluation (average DSC 0.6808).
Predictions on the test set (DSC 0.7367) were obtained by majority voting across the segmentation results of all 5 models.

In~\cite{murugesan2022head}, Murugesan et al. (team ``BAMF Health'') proposed to ensemble the predictions of 3D nnU-Nets (with and without residual connections) using adaptive ensembling to eliminate false positives. A selective ensemble of 8 test-time augmentations and 10 folds (5 U-Nets and 5 residual U-Nets) was used for the final segmentation output.

 In~\cite{wang2022ccut}, Wang et al. (team ``Wangjiao'') used a combination of convolutional and transformer blocks in a U-Net model with attention (global context and channel) in the decoder. The model was trained with squeeze and excitation, and a Dice and Focal loss.

In Table~\ref{tab:algoSegmentation}, we summarize some of the main components of the participants' algorithms, including model architecture, preprocessing, training scheme and postprocessing.

\subsubsection{Results}
The results, including average DSC and HD95 are summarized in Table~\ref{tab:algoSegmentation} with an algorithm summary. The two results at the bottom of the table without a rank were made ineligible to the ranking due to an excessive number of submissions on the HECKTOR 2020 dataset (on the online leaderboard) resulting in an overfit of the 2020 test set which represents half of the 2021 test set.

\begin{figure}
\centering
\begin{subfigure}{.48\textwidth}
  \centering
  \includegraphics[width=0.98\linewidth]{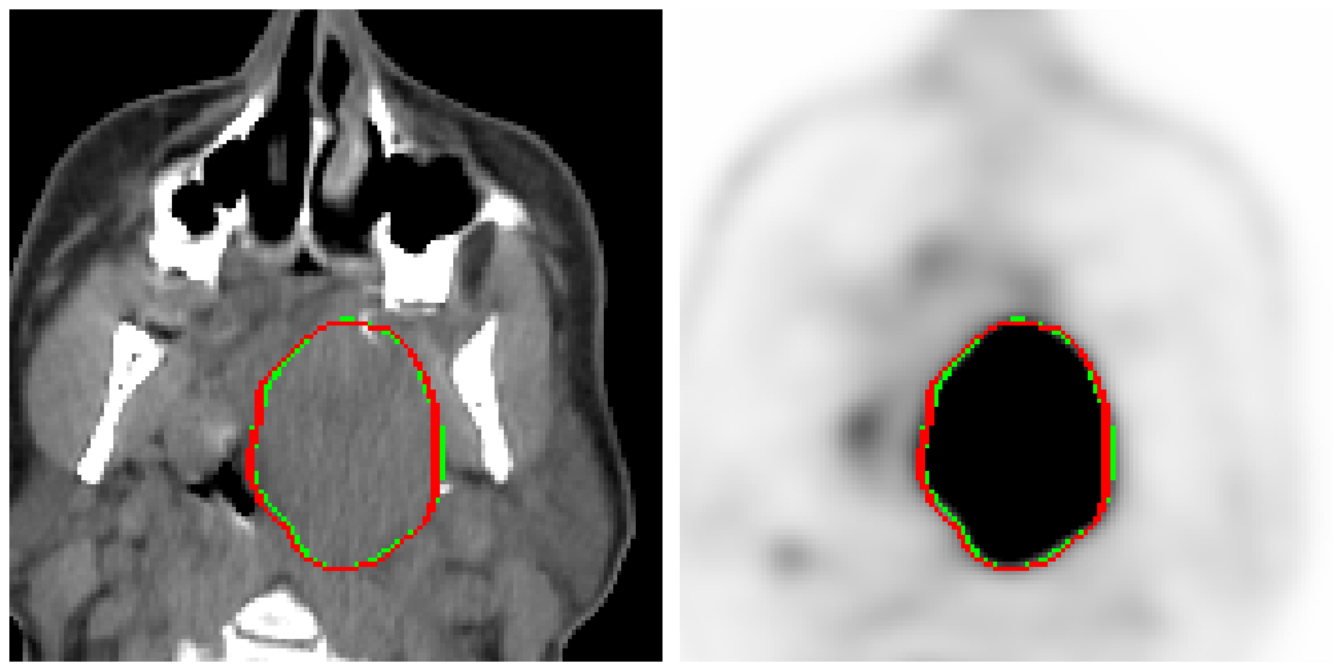}
  \caption{CHUV020, DSC=0.9493}
\end{subfigure}%
\begin{subfigure}{.48\textwidth}
  \centering
  \includegraphics[width=0.98\linewidth]{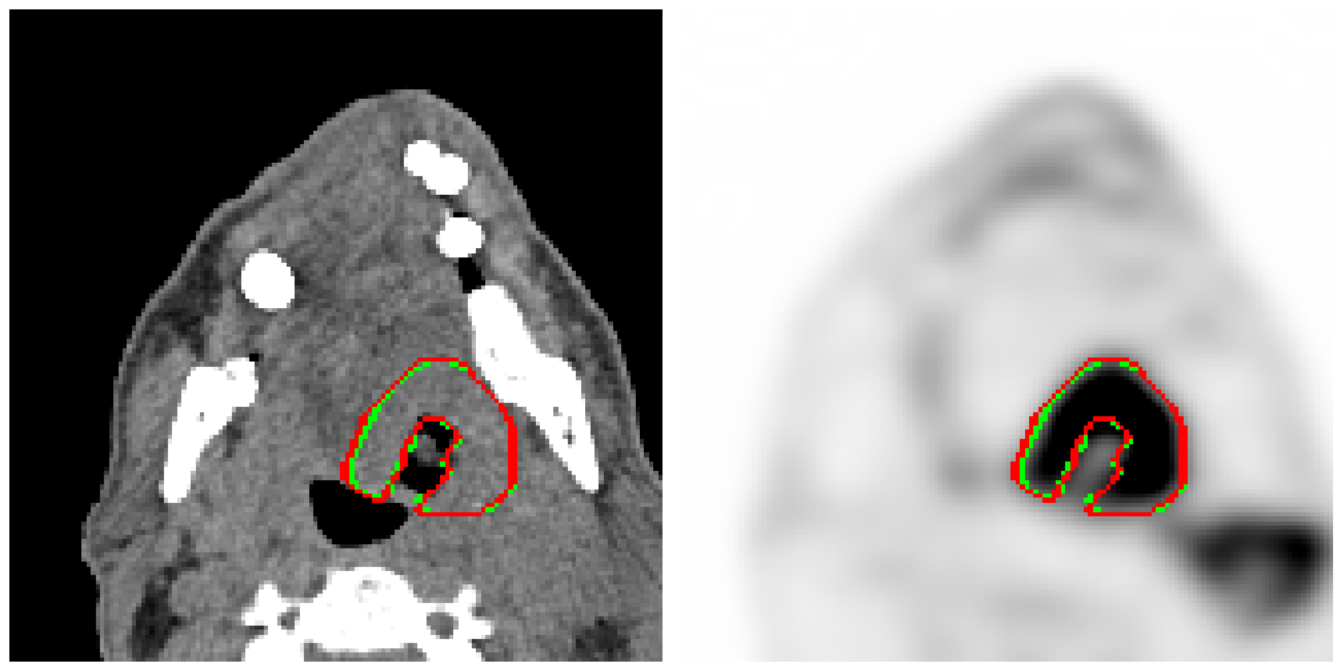}
  \caption{CHUP051, DSC=0.9461}
\end{subfigure}
\begin{subfigure}{.48\textwidth}
  \centering
  \includegraphics[width=0.98\linewidth]{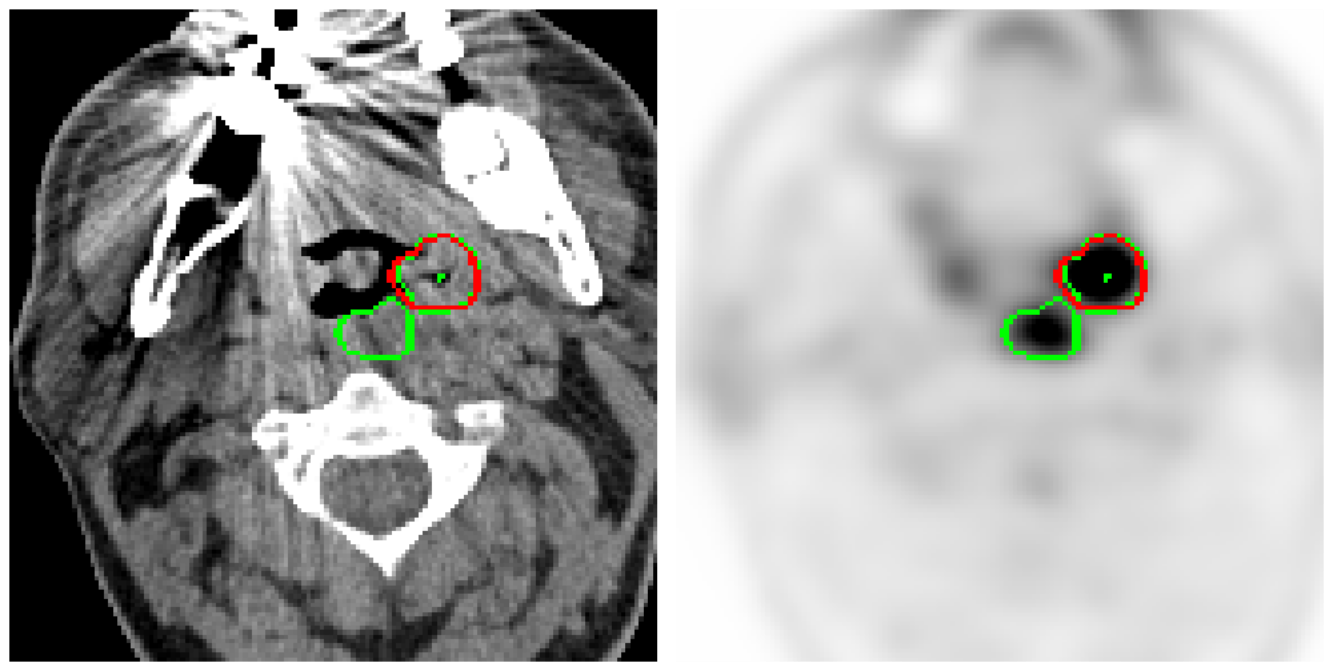}
  \caption{CHUP063, DSC=0.3884} 
\label{fig:examples_results_fp1}
\end{subfigure}
\begin{subfigure}{.48\textwidth}
  \centering
  \includegraphics[width=0.98\linewidth]{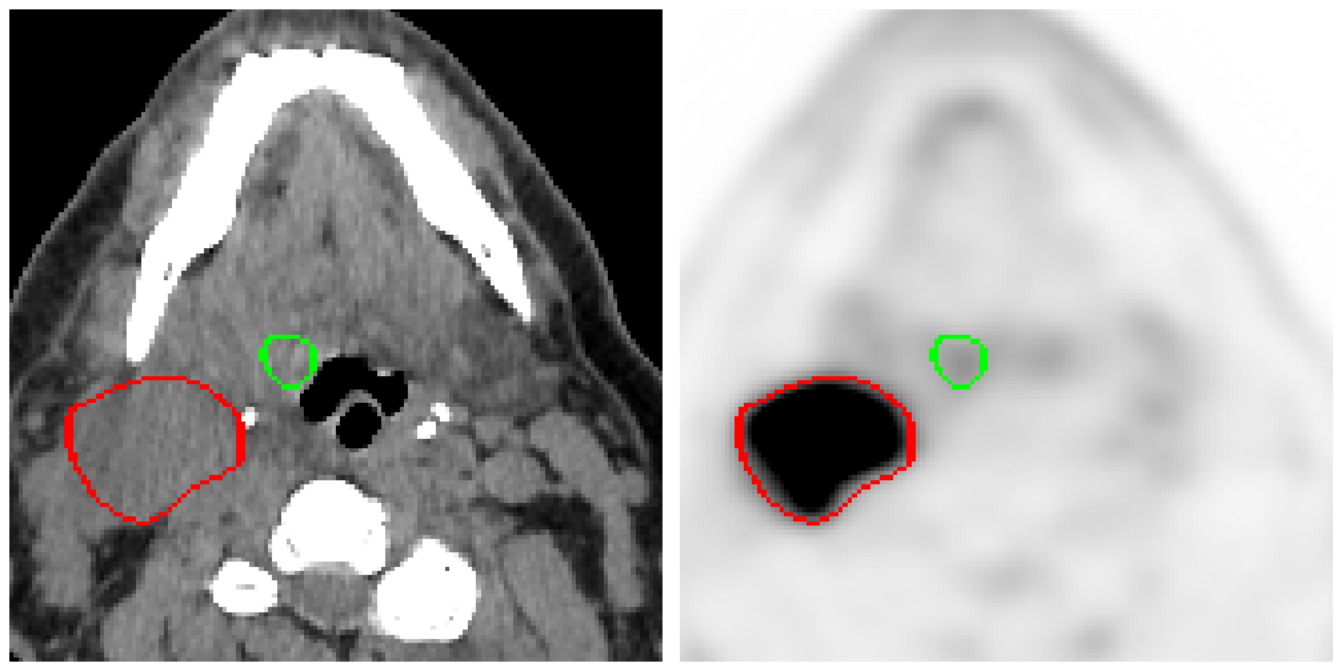}
  \caption{CHUV036, DSC=0.0000} 
  
\label{fig:examples_results_fp2}
\end{subfigure}
\caption{Examples of results of the winning team (Pengy~\cite{xie2022head}). The automatic segmentation results (green) and ground truth annotations (red) are displayed on an overlay of 2D slices of PET (right) and CT (left) images. The reported DSC is computed on the whole image, see Eq.~\eqref{eq:dice_score}. 
}
\label{fig:examples_results}
\end{figure}

The results from the participants range from an average DSC of 0.6331 to 0.7785 and the median HD95  from 6.3718 to 3.0882. Xie and Peng.~\cite{xie2022head} (team ``Pengy'') obtained the best overall results with an average DSC of 0.7785 and a median HD95 of 3.0882. 
Examples of segmentation results (true positives on top row, and false positives on bottom row) are shown in Fig.~\ref{fig:examples_results}.

\section{Tasks 2 and 3: Outcome Prediction}\label{sec:task23}
In order to expand the scope of the challenge compared to the previous installment (2020) that focused on a single task dedicated to the automatic segmentation of GTVt (i.e., same as the updated Task 1 in the 2021 edition), it was decided to add a task with the aim of predicting outcome, i.e. Progression-Free Survival (PFS). 
\subsection{Methods: Reporting of Challenge Design}\label{sec:tasks23-methods}
It was chosen to carry out this task on the same patients dataset used for Task~1, exploiting both the available clinical information and the multimodal FDG-PET/CT images.
The available clinical factors included center, age, gender, TNM 7/8th edition staging and clinical stage, tobacco and alcohol consumption, performance status, HPV status, treatment (radiotherapy only or chemoradiotherapy). The information regarding tobacco and alcohol consumption, performance and HPV status was available only for some patients. For five patients from the training set, the weight was unknown and was set at 75kg to compute SUV values.
Of note, this outcome prediction task was subdivided into two different tasks that participants could choose to tackle separately: Task 3 provided the same data as Task 2, with the exception of providing, in addition, the reference expert contours (i.e., ground-truth of the GTVt). In order to avoid providing the reference contours to participants that could also participate in Task 1, we relied on a Docker-based submission procedure: participants had to encapsulate their algorithm in a Docker and submit it on the challenge platform. The organizers then ran the Dockers on the test data locally, in order to compute the performance. In such a way, the participants never had direct access to the reference contours of the test set, although they could incorporate them in their algorithms the way they saw fit.
\subsubsection{Assessment aim} \mbox{}
The chosen clinical endpoint to predict was PFS. Progression was defined based on Response Evaluation Criteria In Solid Tumors (RECIST) criteria, i.e., either a size increase of known lesions (i.e., change of T and or N), or appearance of new lesions (i.e., change of N and/or M). Disease-specific death was also considered a progression event for patients previously considered stable. In the training set, participants were provided with the survival endpoint to predict, censoring and time-to-event between PET/CT scan and event (in days).

\subsubsection{Assessment Method}
For Task 2, challengers had to submit a CSV file containing the patient IDs with the outputs of the model as a predicted risk score anti-concordant with the PFS in days. For Task 3, the challengers had to submit a Docker encapsulating their method which was run by the organizers on the test set, producing the CSV file for evaluation.
Thus for both tasks, the performance of the output predicted scores were evaluated using the Concordance index (C-index)~\cite{harrell1982evaluating} on the test data. The C-index quantifies the model’s ability to provide an accurate ranking of the survival times based on the computed individual risk scores, generalizing the Area Under the ROC Curve (AUC). It can account for censored data and represents the global assessment of the model discrimination power. Therefore the final ranking was based on the best C-index value obtained on the test set, out of the maximum of 5 submissions per team. The C-index computation is based on the implementation found in the Lifelines library~\cite{davidson2019lifelines} and adapted to handle missing values that are counted as non-concordant. 

\subsection{Results: Reporting of Challenge Outcome}\label{sec:results}

\subsubsection{Participation}\label{sec:participation}
Thirty different teams submitted a total of 149 valid submissions to Task 2. Eighteen corresponding papers were submitted, which made the submissions eligible for final ranking and prize. Probably because of the added complexity of Task 3 requiring encapsulating the method in a Docker, only 8 teams submitted a total of 27 valid submissions. All these 8 teams also participated in Task 2, with 7 corresponding papers.

\subsubsection{Outcome prediction: summary of participants' methods}\label{sec:algoPFS}

The following describes the approach of each team participating in Task 2 (and 3 for some), in the order of the Task 2 ranking. Table \ref{tab:algoPFS} provides a synthetic comparison of the methodological choices and designs for these tasks.

\begin{landscape}
\begin{table}[]
\begin{tabular}{
l|c|cccccc|ccc|cccc|ccccccccccc|c|ccc}
 & {} & \multicolumn{6}{c}{Pre-processing} & \multicolumn{3}{c}{Segment.} & \multicolumn{4}{c}{Image features} & \multicolumn{11}{c}{Modeling and training approach} & {} & \multicolumn{3}{c}{Masks} \\
Team & 
\textbf{\makecell{C-index\\Task 2}} & 
\rotatebox[origin=l]{90}{Iso-resampling} & 
\rotatebox[origin=l]{90}{CT clipping} & 
\rotatebox[origin=l]{90}{Min-max norm.} & 
\rotatebox[origin=l]{90}{Standardization} & 
\rotatebox[origin=l]{90}{PET/CT fusion} & 
\rotatebox[origin=l]{90}{Further cropping} & 
\rotatebox[origin=l]{90}{Relies on Task 1} & 
\rotatebox[origin=l]{90}{Additional segm.} & 
\rotatebox[origin=l]{90}{No segmentation} & 
\rotatebox[origin=l]{90}{Deep features} & 
\rotatebox[origin=l]{90}{Large radiomics set} & 
\rotatebox[origin=l]{90}{Volume, shape} & 
\rotatebox[origin=l]{90}{IBSI compliant} & 
\rotatebox[origin=l]{90}{Ensembling} & 
\rotatebox[origin=l]{90}{Deep model} & 
\rotatebox[origin=l]{90}{Algo. RF, SVM…} & 
\rotatebox[origin=l]{90}{Feature selection} & 
\rotatebox[origin=l]{90}{PET as input} & 
\rotatebox[origin=l]{90}{CT as input} & 
\rotatebox[origin=l]{90}{PET/CT fusion} & 
\rotatebox[origin=l]{90}{Use clinical var.} & 
\rotatebox[origin=l]{90}{Imputed missing} & 
\rotatebox[origin=l]{90}{Cross-val.} & 
\rotatebox[origin=l]{90}{Augmentation} & 
\textbf{\makecell{C-index\\Task 3}} & 
\rotatebox[origin=l]{90}{GT masks} &
\rotatebox[origin=l]{90}{Task 1 masks} &
\rotatebox[origin=l]{90}{PET thresh. masks} \\
\hline
BioMedIA \cite{saeed2022ensemble} & 0.7196 &  &  & { \checkmark} &  & { \checkmark} & {{ \checkmark}} & {} &  & { \checkmark} & { \checkmark} & {} & {} & {} & {} & { \checkmark} & {} & {} & {} & {} & {{ \checkmark}} &{ \checkmark} &  & {} & {} & na &  &  &  \\
\hline
Fuller MDA \cite{naser2022progression} & 0.6938 &  & { \checkmark} & { \checkmark} & { \checkmark} & {} &  & {} &  & { \checkmark} & { \checkmark} & {} & {} & {} & { \checkmark} & { \checkmark} & {} & {} & { \checkmark} & { \checkmark} & & { \checkmark} &  & {} & { \checkmark} & 0.6978 & \checkmark & { \checkmark} &  \\
\hline
Qurit Tecvico \cite{paeenafrakati2022advanced}& 0.6828 &  & {} & { \checkmark} & {} & {} &  & { \checkmark} & & {} & {} & { \checkmark} & {} & { \checkmark} & { \checkmark} & {} & { \checkmark} & { \checkmark} & {} & {} & {{ \checkmark}} & { \checkmark} &  & { \checkmark} & {} & na &  &  &  \\
\hline
BMIT\_USYD \cite{meng2022multi}& 0.6710 &  & { \checkmark} & { \checkmark} & { \checkmark} &  & {{ \checkmark}} & { \checkmark} & & {} & { \checkmark} & {} & {} & {} & { \checkmark} & { \checkmark} & {} & {} & { \checkmark} & { \checkmark} &  & { \checkmark} &  & {} & { \checkmark} & na &  &  &  \\
\hline
DMLang \cite{lang2022deep}& 0.6681 &  & { \checkmark} & { \checkmark} &  &  &  & { \checkmark} & {} &  & { \checkmark} &  &  &  &  & { \checkmark} &  &  & { \checkmark} & { \checkmark} & {} & { \checkmark} & {} &  & { \checkmark} & na &  &  &  \\
\hline
TECVICO\_C. \cite{fatan2022fusion}& 0.6608 &  &  &  &  & { \checkmark} & {{ \checkmark}} & { \checkmark} & {} &  &  & { \checkmark} &  & { \checkmark} &  &  & { \checkmark} & { \checkmark} &  &  & {{ \checkmark}} & { \checkmark} & {} & { \checkmark} &  & na &  &  &  \\
\hline
BAMF Health \cite{murugesan2022head}& 0.6602 &  & { \checkmark} & { \checkmark} & { \checkmark} &  &  & { \checkmark} & { \checkmark} &  &  & { \checkmark} &  & { \checkmark} & {} & {} & { \checkmark} & { \checkmark} & { \checkmark} & { \checkmark} & {} & { \checkmark} & {{ \checkmark}} & { \checkmark} &  & 0.6602 &  &  & { \checkmark} \\
\hline
ia-h-ai \cite{starke2022hybrid}& 0.6592 &  &  &  &  &  &  &  &  &  &  & { \checkmark} &  & { \checkmark} & {} & {} & { \checkmark} & { \checkmark} & { \checkmark} & { \checkmark} &  & { \checkmark} &  & { \checkmark} & {} & 0.6592 &  &  & { \checkmark} \\
\hline
Neurophet \cite{lee2022dual}& 0.6495 &  &  &  &  &  &  & { \checkmark} & {} &  &  &  & { \checkmark} & {} & {} & {} & { \checkmark} & {} & {} & {} &  & { \checkmark} &  & { \checkmark} & {} & na &  &  &  \\
\hline
UMCG \cite{ma2022self}& 0.6445 &  & { \checkmark} & { \checkmark} & { \checkmark} & { \checkmark} &  & & {} & { \checkmark} & { \checkmark} & {} & {} & {} & { \checkmark} & { \checkmark} & {} & {} & { \checkmark} & { \checkmark} & {{ \checkmark}} & { \checkmark} &  & { \checkmark} & { \checkmark} & 0.6373 & \checkmark & { \checkmark} &  \\
\hline
Aarhus Oslo \cite{ren2022comparing}& 0.6391 &  &  &  &  &  &  &  & {} &  &  &  &  &  &  &  &  &  &  &  & {} & { \checkmark} &  & { \checkmark} &  & na &  &  &  \\
\hline
RedNeucon \cite{asenjo2022pet}& 0.6280 &  &  &  &  &  &  & { \checkmark} & {{ \checkmark}} &  &  & { \checkmark} &  &  &  &  & { \checkmark} & { \checkmark} & {} & {} &  & { \checkmark} &  & { \checkmark} & {} & na &  &  &  \\
\hline
Emmanuelle B. \cite{bourigault2022pet}& 0.6223 & {{ \checkmark}} & { \checkmark} & {} & { \checkmark} & {} &  & { \checkmark} &  & {} & { \checkmark} & { \checkmark} & {} & { \checkmark} & {} & { \checkmark} & { \checkmark} & { \checkmark} & { \checkmark} & { \checkmark} &  & { \checkmark} & {{ \checkmark}} & { \checkmark} & {} & na &  &  &  \\
\hline
BCIOQurit & 0.6116 \cite{yousefirizi2022segmentation}& {\checkmark} & \checkmark & {} & \checkmark & {} &  & { \checkmark} &  & {} & { \checkmark} & { \checkmark} & {} & { \checkmark} & {} & {} & { \checkmark} & { \checkmark} & { \checkmark} & { \checkmark} &  & { \checkmark} &  & { \checkmark} & { \checkmark} & 0.4903 & { \checkmark} &  &  \\
\hline
Vokyj \cite{muller2022deep}& 0.5937 & {{ \checkmark}} & { \checkmark} & {} & { \checkmark} & {} &  & { \checkmark} &  & {} & {} & {} & { \checkmark} & {} & {} & {} & { \checkmark} & {} & {} & {} &  & { \checkmark} &  & {} & { \checkmark} & na &  &  &  \\
\hline
Xuefeng \cite{ghimire2022head}& 0.5510 & {{ \checkmark}} & { \checkmark} & {} & { \checkmark} & {} &  & { \checkmark} &  & {} & {} & {} & { \checkmark} & {} & {} & {} & { \checkmark} & {} & {} & {} &  & {} &  & {} & {} & 0.5089 & { \checkmark} &  &  \\
\hline
DeepX \cite{yuan2022automatic}& 0.5290 & {{ \checkmark}} & { \checkmark} & {} & { \checkmark} & {} &  & { \checkmark} &  & {} & {} & { \checkmark} & {} & { \checkmark} & {} & {} & { \checkmark} & {} & { \checkmark} & { \checkmark} &  & { \checkmark} &  & {} & {} & na &  &  & \\
\hline
\end{tabular}
\caption{Synthetic comparison of outcome prediction methods. More details are available in Section~\ref{sec:algoPFS}. All participants of task 3 also participated in task 2.}\label{tab:algoPFS}
\end{table}
\end{landscape}

In~\cite{saeed2022ensemble}, Saeed et al. (team ``BiomedIA'') first experimented with the clinical variables and determined that better prediction was achieved using only variables with complete values, compared to using all variables with imputing missing values. They elected to first implement a fusion of PET and CT images by averaging them into a new single PET/CT image that would be further cropped (2 different sizes of 50x50x50 and 80x80x80 were tested) to form the main input to their solution based on a 3D CNN (Deep-CR) trained to extract features which were then fed into Multi-Task Logistic Regression (MTLR, a sequence of logistic regression models created at various timelines to evaluate the probability of the event happening) improved by integrating neural networks to achieve nonlinearity, along with the clinical variables. Two different models were compared as the input to MTLR: either a CNN with 3 paths (for PET, CT and fused PET/CT) or only one using only fused PET/CT. The batch size, learning rate, and dropout were experimentally set to 16, 0.016, and 0.2 respectively for the training. The model was trained for 100 epochs using Adam optimizer. No cross-validation or data augmentation was used. Of note, the results of CNN and MTLR (i.e., exploiting both images and clinical variables) were averaged with the prediction of a Cox model using only clinical variables to obtain the best result. This team won Task 2 with a C-index of 0.72 but did not participate in Task 3.

In~\cite{naser2022progression}, Naser et al. (team ``Fuller MDA'')  also adopted an approach based on deep learning. Clinical variables without missing values were transformed into an image matrix in order to be fed along with PET and CT images (rescaled and z-score normalized) as separate channels to a DenseNet121 CNN. Adopting a 10-fold cross-validation scheme, the model was trained either only with 2 channels (PET and CT) or 3 (adding the clinical), with data augmentation, for 800 iterations with a decreasing learning rate, the Adam optimizer and a negative log-likelihood loss. Of note, the PFS was discretized into 20 discrete intervals for the output of the network. Two different approaches of ensembling the various models obtained over the 10 folds (consensus or average) were implemented. The best result (0.694, rank 2) was obtained with the Image+Clinical consensus model. The team also participated in Task 3 where they used ground-truth masks as an additional input channel to the same network \cite{wahid2022combining}, achieving the first rank with a C-index of 0.70.

In~\cite{paeenafrakati2022advanced}, Paeenafrakati et al. (team ``Qurit Tecvico'') implemented a classical radiomics approach, where a large set of IBSI-compliant features were extracted with the SERA package~\cite{ashrafinia2019quantitative} from the delineated tumor (based on the output of their solution for Task 1) in PET, CT as well as a fusion of PET/CT (of note, 10 different fusion techniques were explored). The features were then selected through 13 different dimensionality reduction techniques and 15 different selection methods and combined along with clinical variables, into several models with 5-fold cross-validation (the entire training set was used for each approach) through the use of 8 different survival prediction algorithms. The best performance (0.68) in the test set was obtained with an ensemble voting of these various algorithms, obtaining third rank in Task 2 (the team did not participate in Task 3).

In~\cite{meng2022multi}, Meng et al. (team ``BMIT USYD'') proposed a single unified framework to achieve both segmentation (Task 1) and outcome prediction (Task 2, no participation in Task 3). They first selected a few relevant clinical variables to take into account by performing a univariate/multivariate analysis, retaining only HPV status, performance status, and M stage. Their proposed model is composed of two main components: a U-Net based network for segmentation and a DenseNet based cascaded survival network. Both extract deep features that are fed into fully connected layers for outcome prediction and are trained in an end-to-end manner to minimize the combined loss of segmentation and survival prediction losses, with Adam optimizer, a batch size of 8 for 10000 iterations, with a decreasing learning rate. Clinical factors were concatenated in the non-activated fully connected layer. Of note, both the segmentation output and the cropped, normalized PET and CT images are fed to the DenseNet cascaded survival network. Data augmentation (random translations, rotations and flipping) was applied. Ten different models were trained, the first 5 through leave-one-center-out cross-validation and the next five with 5-fold cross-validation. The ensemble of these achieved a C-index of 0.671 in the test set.

In~\cite{lang2022deep}, Lang et al. (team ``DMlang'') relied on the segmentation output of Task 1 (or on the reference contours in training) to generate cropped bounding-boxes as inputs to their approach for predicting outcome, which relied on extracting deep features from PET and CT images thanks to a pre-trained C3D network designed to classify video clips. In order to feed PET and CT images to this C3D model, each 3 consecutive slices were fed to the color channels. The obtained PET and CT features were then concatenated and fed to a dense layer, which was then concatenated with clinical variables. Each output neuron represented the conditional probability of surviving a discrete time interval (the best model involved layers of size 512 and 256 and an output size of 15 corresponding to time intervals covering a maximum of 10 years of survival with the first 5 years split into intervals of half a year and all subsequent intervals with a width of one year). The same data augmentation as for the segmentation task was used. For training this network, a batch size of 16 was applied and 75 epochs were used with the Adam optimizer to minimize the negative log-likelihood. For model optimization, hyper-parameters were tuned manually. Of note, the team did not rely on ensemble of models nor on cross-validation, but generated a single stratified split of the training data. The trained model achieved a C-index of 0.668. The team did not participate in Task 3.

In~\cite{fatan2022fusion}, Fatan et al. (team ``TECVICO Corp Family'') used a similar PET/CT fusion approach (5 different techniques) and cropping as the team "Qurit\_Tecvivo", extracted 215 IBSI-compliant radiomics features with the same package (SERA), that were fed into a number of feature selection techniques (7) and classifiers (5). They did not perform an ensemble of these but selected the best model in cross-validation during training. The best combination (LP-SR fusion and the classifier GlmBoost) obtained 0.66 in the test set. They did not participate in Task 3.

In~\cite{murugesan2022head}, Murugesan et al. (team ``BAMF Health'') participated in both Tasks 2 and 3. Interestingly, their best results were obtained using the tumor masks by their segmentation method of Task 1, instead of the reference contours. Their solution was based on standard IBSI-compliant radiomics features extracted with Pyradiomics from PET and CT images after z-score normalization of intensities. In addition, in-house features calculating the number of uptakes and their volumes in each PET/CT were calculated through thresholding of PET SUVs. All clinical variables were exploited, missing values were imputed using the mean value of provided variables. Before further exploitation of the radiomics features, they were standardized using their mean and standard deviation. Then principal component analysis was applied to the features, capturing 95 of information. Variable importance combined with fast unified random forests for survival, regression, and classification was used for modeling through repeated random sub-sampling validation over 100 multiple random splits, in order to look for an optimal combination of features and to optimize hyper-parameters. The best result in the test set was obtained with a model relying on PCA components, with a 0.66 C-index (for both Tasks 2 and 3).

In~\cite{starke2022hybrid}, Starke et al. (team ``ia-h-ai'') built a strategy based on standard radiomics modeling, addressing both Tasks 2 and 3. They first strategically split the training data into 3 folds, ensuring that for each split, one of the centers is present in the validation set but not the training. Clinical factors were all considered, by imputing missing values through k-nearest neighbor (k=20). They used either the provided reference volumes or alternative ones obtained through thresholding the PET intensities with SUV $>$ 2.5. 172 IBSI-compliant handcrafted features were then extracted from both PET and CT images volumes of interest using Pyradiomics. They established some baseline models through Cox proportional hazards models exploiting only the clinical variables, then moved to more advanced modeling relying on random survival forest, still using only clinical variables. In order to add image features to the models, they performed feature selection through three different processes: stability (L1-regularized Cox regression applied to multiple bootstrapped datasets for a range of regularization strength parameters), permutation-based feature importance and finally sequential feature selection. This allowed them to retain only a small number of features for the actual modeling step, where they compared different approaches using random forest survival (300 trees): fully automated feature selection and combination or different ways of manually selecting features, including a priori selection based on literature. They consistently obtained better performance on the test set by relying on the alternative volumes of interest (thresholded at SUV $>$ 2.5, leading to volumes larger than the reference ground-truth contours), and models with hand-picked features, contrary to fully automatic selection that demonstrated overfitting.

In~\cite{lee2022dual}, Lee et al. (team ``Neurophet'') exploited only clinical variables (missing values were coded as 0 or -1 depending on the variable) and segmented volumes from Task 1 (i.e. only 1 feature, the tumor volume) to train a random forest survival model through 5-fold randomized cross-validation with 100 iterations. Of note, the center ID was added as a clinical factor. The proposed model achieved a C-index of 0.65 on the test set, with a higher performance than the same model without tumor volume (0.64). The team did not participate in Task~3.

In~\cite{ma2022self}, Ma et al. (team ``UMCG'') proposed a pipeline based on deep learning as well, consisting of three parts: 1) the pyramid autoencoder of a 3D Resnet extracting image features from both CT and PET, 2) a feed-forward feature selection to remove the redundant image and clinical features, and 3) a DeepSurv (a Cox deep network) for survival prediction. Clinical variables were used but missing values were not imputed, rather described as an additional class (i.e., unknown). PET and CT images were pre-processed and a new PET/CT image obtained by summation of PET and CT was used as a third input to the autoencoder. The segmentation masks were not used for Task 2, but were used for Task 3 in order to extract the tumor region in two different ways, both being used as inputs to the network. This pipeline was trained on using different splits of the training set (leave-one-center out and random selection of 179 patients for training and 45 for validations), resulting in 6-fold cross-validation. The Autoencoders were trained using the Adam optimizer with the initial learning rate 0.001 and data augmentation for 80 epochs. The official DeepSurv was trained for 5000 steps with the default settings. A total of 30 DeepSurv models were trained in each fold and the 3 models with the highest validation set C-index were selected. In total 18 models were obtained and their predicted risk scores are averaged to obtain the final result: 0.6445 and 0.6373 C-index in the test set for Task 2 and 3 respectively.

In~\cite{ren2022comparing}, Ren et al. (team ``Aarhus Oslo'') team compared a conventional radiomics approach (although without tumor delineation, i.e., features were extracted from the whole bounding-box) and a deep learning approach in Task 2 only. Both used the provided bounding-box of PET and CT images as inputs, and in the case of the deep learning approach, an additional pre-processing step was applied to PET images in order to reduce the variability of images due to various centers based on a sin transform. For the standard radiomics approach, only clinical variables without missing values were exploited, whereas they were not used in the case of the deep learning approach. In the standard radiomics modeling, over 100 IBSI-compliant features were calculated but only a handful were manually selected based on literature and further used: one from CT and 4 from PET. These features (and clinical variables) were then fed to 2 ensemble models: random forest and gradient boosting. Hyper-parameters (number of trees, maximum depth for each tree, and learning rate, loss function tuning) were tuned using grid-search, and models were trained and evaluated using 5-fold cross-validation. In the case of deep learning, only CT and PET-sin images were used as input of a CNNs built with the encoder part of the SE Norm U-Net model~\cite{iantsen2021squeeze} with three fully connected layers (4096, 512, and 1 units) added to the top. Five-fold cross-validation was also used. Each model was trained for 150 epochs using the Adam optimizer with a batch size of 4. The initial learning rate was set to 3e-6 and the loss was defined as a fusion of the Canberra distance loss and Huber loss ($\delta = 1$). Based on the results of cross-validation in training, the four following models were evaluated on the test set: Gradient boosting trained on either clinical factors (either all or only uncensored data) or both clinical factors and selected radiomics features and ensemble based on mean predicted values of five-fold deep learning models trained on FDG-PET/CT. All models had near-random performance in the test set, except the clinical-only model built with gradient boosting (0.66).

In~\cite{asenjo2022pet}, Asenjo et al. (team ``RedNeucon'') implemented a conventional radiomics approach based on the extraction of handcrafted features from PET and CT with a Matlab toolbox, from the reference contour volumes and the segmentation output of Task 1, as well as an additional volume of interest generated by determining a two pixel inward and outward the contours to get a tumor “boundary region”. Only clinical variables without missing values were used. Features were then selected after ranking according to 2 methods, ranking for classification using a Fisher F-Test and an algorithm based on K-nearest neighbors. When two features showed a correlation above 0.5, the best one was kept. Three different modeling algorithms were compared in 5-fold cross-validation: Gaussian Process Regression (GPR), an Ensembled Bagged of trees and a Support Vector Machine. The best result on the test set (0.628) was obtained with the GPR with 35 features.

In~\cite{bourigault2022pet}, Bourigault et al. (team ``Emmanuelle Bourigault'') proposed a Cox proportional hazard regression model using a combination of clinical, radiomic, and deep learning features from PET/CT images. All clinical variables were exploited, after imputing missing values using a function of available ones. IBSI-compliant handcrafted radiomics features including wavelet-filtered ones were calculated using Pyradiomics and were combined with deep features from the 3D U-Net used in the segmentation Task 1, in addition to clinical variables. Spearman rank correlation above 0.8 was used to eliminate intercorrelated features. Feature selection was performed using Lasso regression with 5-fold cross-validation, reducing the set of 270 variables to 70 (7 clinical, 14 radiomics and 49 deep). Three different models were implemented for modeling: Cox proportional hazard regression model, random survival forest and Deepsurv (a Cox proportional hazards deep neural network). All three models were trained with different combinations of the selected clinical, radiomics (PET, CT or PET/CT) and deep features. The best performance in validation was obtained with the Cox model using clinical + CT radiomics + deep learning features, although in the test set its final performance was 0.62.

In~\cite{yousefirizi2022segmentation}, Yousefirizi et al. (team ``BCIOqurit'') proposed training a proportional hazard Cox model with a multilayer perceptron neural net backbone to predict the score for each patient. This Cox model was trained on a number of PET and CT radiomics features extracted from the segmented lesions, patient demographics, and encoder features provided from the penultimate layer of a multi-input 2D PET/CT convolutional neural network tasked with predicting time-to-event for each lesion. A grid search over several feature selection and classifiers methods identified 192 unique combinations of radiomics features that were used to train the overall Cox model with the Adam optimizer, a learning rate of 0.0024, a batch size of 32, and an early stopping method that monitored the validation loss. A 10-fold cross-validation scheme was used and an ensemble model of these achieved a C-index score of 0.612 in the test set.

In~\cite{muller2022deep}, Muller et al. (team ``Vokyj'') proposed to fit a Weibull accelerated failure time model with clinical factors and the shape descriptors of the segmented tumor (output of Task 1). M-stage and two shape features (Euler number and Surface Area) were the most predictive of PFS, the model achieving a performance of 0.59 in the test set. The team did not participate in Task 3.

In~\cite{ghimire2022head}, Ghimire et al. (team ``Xuefeng'') implemented a straightforward approach that consisted in calculating the tumor volume and tumor surface area of the Task 1 segmentation outputs, as well as the classification output from the segmentation network trained to classify the input images into 6 different classes of PFS (which was first discretized into 6 bins). These imaging features were then combined with all available clinical factors, for which missing values were imputed with the median value for numerical variables and mode value for categorical ones. All features were then normalized to zero mean and 1 standard deviation for a linear model to be fitted to the training data. The model was applied to both Tasks 2 and 3, using the reference contours instead of the Task 1 segmentation results, leading to C-index values of ~0.43 and ~0.51 respectively.

In~\cite{yuan2022automatic}, Yuan et al. (team ``DeepX'') implemented a standard radiomics approach, extracting more than 200 IBSI-compliant handcrafted features with Pyradiomics, from both PET and CT images using the segmentation output of Task 1, which were then manually ranked and selected according to their concordance index. Regarding clinical variables, only age was used. The 7 selected features were evaluated independently or combined through averaging concordance ranking, obtaining their best C-index of 0.53 in the test set.

\section{Discussion: Putting the Results into Context}\label{sec:discussion}

Outcomes and findings of participating methods are summarized in Section~\ref{sec:outcomes} for all three tasks.
In general, we observed that simplicity was beneficial for generalization and that sophisticated methods tend to overfit the training/validation.
Despite the diversity in terms of centers and image acquisition, no specific feature or image harmonization method was employed, which could be one avenue for improving generalization abilities of the methods in all tasks~\cite{AIW2021}.

The combined scope of the three proposed tasks also allowed the emergence of very interesting findings concerning the relationship of the GTVt contouring task and PFS prediction. 
In a nutshell, ground truth ROIs were not providing top results, even though they were re-annotated in a centralized fashion to be dedicated for radiomics~\cite{FAO2021b}.
Simple PET thresholded and bounding-boxes for deep learning outperformed the use of ground truth ROI.
This suggests that algorithms looking elsewhere than the GTVt is beneficial (e.g. tumoral environment, nodal metastases).
Fully automatic algorithms are expected to provide optimal results, which was already highlighted by several papers in the context of the HECKTOR challenge~\cite{fontaine2021fully,starke2022hybrid,andrearczyk2021multi,naser2022progression,murugesan2022head}.
This potentially obviates the need for GTVt contouring, opening avenues for reproducible and large scale radiomics studies including thousands potential subjects.

\subsection{Outcomes and Findings}\label{sec:outcomes}
A major benefit of this challenge is to compare various algorithms developed by teams from all around the world on the same dataset and task, with held-out test data. 

We distinguish here between the technical and biomedical impact.
The main technical impact of the challenge is the comparison of state-of-the-art algorithms on the provided data. We identified key elements for addressing the task: 3D U-Net, preprocessing, normalization, data augmentation and ensembling, as summarized in Tables~\ref{tab:algoSegmentation} and~\ref{tab:algoPFS}.  
The main biomedical impact of the results is the opportunity to generate large cohorts with automatic tumor segmentation for comprehensive radiomics studies, as well as to define and further push state of the art performance. 

\paragraph{Task 1: Automatic segmentation of the GTVt}\mbox{}
\\
The best methods obtain excellent results with DSCs above $0.75$, better than inter-observer variability (DSC 0.61) performed on a subset of our data and similar variability reported in the literature (DSCs of 0.57 and 0.69 on CT and PET/CT respectively)~\cite{gudi2017interobserver}. Note that without injected contrast CT, delineating the exact contour of the tumor is very difficult. Thus, the inter-observer DSC could be low only due to disagreements at the border of the tumor, without taking into account the error rate due to the segmentation of non-malignant structures (if any). For that reason, defining the task as solved solely based on the DSC is not sufficient.
In the context of this challenge, we can therefore define the task as solved if the algorithms follow these three criteria:
\begin{enumerate}
    \item Higher or similar DSC than inter-observers agreement.
    \item Detect all the primary tumors in the oropharynx region (i.e. segmentation not evaluated at the pixel level, rather at the occurrence level).
    \item Similarly, detect only the primary tumors in the oropharynx region (discarding lymph nodes and other potentially false positives).
\end{enumerate}
According to these criteria, the task is partially solved. The first criterion, evaluating the segmentation at the pixel level, is fulfilled. At the occurrence level (criteria 2 and 3), however, even the algorithms with the highest DSC output FP and FN regions. 
Besides, there is still a lot of work to do on highly related tasks, including the segmentation of lymph nodes, the development of super-annotator ground truth as well as the agreement of multiple annotators, and, finally, the prediction of patient outcome following the tumor segmentation.

Similarly to last year's challenge, we identified the same key elements that cause the algorithms to fail in poorly segmented cases. These elements are as follows; low FDG uptake on PET, primary tumor that looks like a lymph node, abnormal uptake in the tongue and tumor present at the border of the oropharynx region. Some examples are illustrated in Fig.~\ref{fig:examples_cases}.  

\paragraph{Tasks 2 and 3: Predicting PFS}\mbox{}\\
The challengers relied on a variety of approaches and tackled the task quite differently (Table \ref{tab:algoPFS}). A few teams relied on deep learning exclusively, whereas others exploited more classical radiomics pipelines. Some teams also implemented various combinations of both. PET and CT images were also exploited in several different ways. Either as separate inputs or through various fusion techniques, for either deep learning or classical radiomics analysis. Interestingly, despite the recent rise of interest in the development of methods dedicated to the harmonization of multicentric data, either in the image domain through image processing or deep learning based image synthesis \cite{cnnkernels} or in the features domain through batch-harmonization techniques such as ComBat \cite{ronrickCombat}, none of the teams implemented specific multicentric harmonization techniques, beyond usual approaches to take into account the diversity of the images in the training and testing sets by relying on, for example, leave-one-center-out cross-validation and image intensities rescaling or z-score normalization. The use of clinical variables was also the opportunity for challengers to deploy different approaches. Among the methods using deep networks, some encoded the clinical information into images to feed them as input to the deep networks, whereas others integrated them as vectors concatenated in other layers. Some teams elected to rely only on clinical factors without missing values, whereas others implemented some way of imputing missing values in order to exploit all available variables. In addition, some teams pre-selected only a subset of the clinical variables with prior knowledge. Interestingly, some challengers obtained their best performance by building models relying only on clinical variables. Finally, most teams who participated in Task 1 relied on their segmentation output in Tasks 2 and 3, however, a few explored additional or alternative volumes of interest. Interestingly for Task 3, some challengers obtained better results using alternative segmentation or Task 1 outputs instead of the provided reference contours.

\paragraph{Predicting PFS was the objective of both Tasks 2 and 3}\mbox{}\\
The only difference was that the GTVt ROI was provided for Task 3, but not for Task 2.
One surprising trend showed that the predictive performance was found to be slightly higher when the GTVt ROI was not used (Task 2), which could be the result of the following. 
First, fewer teams participated in Task 3, which can be partially explained by the requirement to submit a Docker container instead of direct prediction of hazard scores.
Second, for deep learning-based radiomics, using input ROIs is less straightforward than handcrafted radiomics, which makes input contour less relevant. 
Nevertheless, Starke et al.~\cite{starke2022hybrid} used a classical radiomics pipeline and observed that ROIs based on a simple PET-based thresholding approach systematically outperformed a model based on features extracted from the provided GTVt.
This suggests that prognostically relevant information is contained not only in the primary tumor area, but also in other (metabolically active) parts such as the lymph nodes. Similar results have been obtained recently in different tumor localizations. For instance, it was shown in uterine cancer that radiomics features extracted from the entire uterus organ in MRI rather than the tumor only led to more accurate models \cite{uterineseg}. In cervical cancer patients, specific SUV thresholds in PET images led to more accurate metrics \cite{lesueurthresholds}, even though this threshold might not be the more accurate to delineate the metabolic uptake tumor volume. Finally, a study in non-small cell lung cancer recently showed that radiomics features extracted from a large volume of interest containing the primary tumor and the surrounding healthy tissues in PET/CT images could be used to train models as accurate as those trained on features extracted from the delineate tumor, provided a consensus of several machine learning algorithms is used for the prediction \cite{roughsegPETCT}.

\subsection{Limitations of the Challenge}
The dataset provided in this challenge suffers from several limitations. 
First, the contours were mainly drawn based on the PET/CT fusion which is not sufficient to clearly delineate the tumor. Other methods such as MRI with gadolinium or contrast CT are the gold standard to obtain the true contours for radiation oncology. Since the target clinical application is radiomics, however, the precision of the contours is not as important as for radiotherapy planning. 

Another limitation comes from the definition of the task, given that only one segmentation was drawn on the fusion of PET and CT. For radiomics analysis, it could be beneficial to consider one segmentation per modality since the PET signal is often not contained in the fusion-based segmentation due to the poor spatial resolution of this modality.

\section{Conclusions}\label{sec:conclusion}
This paper presented a general overview of the HECKTOR challenge including the data, the participation, main results and discussions. The proposed tasks were the segmentation of the primary tumor in oropharyngeal cancer as well as the PFS prediction. 
The participation was high, with 20, 17, and 6 eligible teams for tasks 1, 2, and 3, respectively.
The participation doubled compared to the previous edition, which shows the growing interest in automatic lesion segmentation for H\&N cancer.

The task proposed this year was to segment the primary tumor in PET/CT images. This task is not as simple as thresholding the PET image since we target only the primary tumor and the region covered by high PET activation is often too large, going beyond the limits of the tumor tissues. Deep learning methods based on U-Net models were mostly used in the challenge. Interesting ideas were implemented to combine PET and CT complementary information. Model ensembling, as well as data preprocessing and augmentation, seem to have played an important role in achieving top-ranking results.

\section*{Acknowledgments}\label{sec:acknowledgments}
The organizers thank all the teams for their participation and valuable work. This challenge and the winner prizes were sponsored by Siemens Healthineers Switzerland, Bioemtech Greece and Aquilab France (500€ each, for Task 1, 2 and 3 respectively). 
The software used to centralise the quality control of the GTVt regions was MIM (MIM software Inc., Cleveland,OH), which kindly supported the challenge via free licences.
This work was also partially supported by the Swiss National Science Foundation (SNSF, grant 205320\_179069) and the Swiss Personalized Health Network (SPHN, via the IMAGINE and QA4IQI projects). 

\newpage
\appendix

\renewcommand{\thesection}{\arabic{section}}

\makeatletter
\renewcommand{\@seccntformat}[1]{Appendix~\csname the#1\endcsname:\quad}
\makeatother

\setcounter{section}{0}

\section{Challenge Information}\label{app:challenge_info}
In this appendix, we list important information about the challenge as suggested in the BIAS guidelines~\cite{maier2020bias}.

\paragraph{Challenge name} \mbox{}
\\HEad and neCK TumOR segmentation and outcome prediction challenge (HECKTOR) 2021
\paragraph{Organizing team} \mbox{}
\\(Authors of this paper) Vincent Andrearczyk, Valentin Oreiller, Sarah Boughdad, Catherine Cheze Le Rest, Hesham Elhalawani, Mario Jreige, John O. Prior, Martin Vallières, Dimitris Visvikis, Mathieu Hatt and Adrien Depeursinge

\paragraph{Life cycle type} \mbox{}
\\A fixed submission deadline was set for the challenge results. 

\paragraph{Challenge venue and platform}\mbox{}
\\The challenge is associated with MICCAI 2021.
Information on the challenge is available on the website, together with the link to download the data, the submission platform and the leaderboard\footnote{\url{www.aicrowd.com/challenges/hecktor}}.

\paragraph{Participation policies}\mbox{}
\\(a) Task 1: Algorithms producing fully-automatic segmentation of the test cases were allowed.
Task 2 and 3: Algorithms producing fully-automatic PFS risk score prediction of the test cases were allowed.
\\(b) The data used to train algorithms was not restricted. If using external data (private or public), participants were asked to also report results using only the HECKTOR data.
\\(c) Members of the organizers' institutes could participate in the challenge but were not eligible for awards.
\\(d) Task 1: The award was 500 euros, sponsored by Siemens Healthineers Switzerland.
Task 2: The award was 500 euros, sponsored by Aquilab.
Task 3: The award was 500 euros, sponsored by Bioemtech.
\\(e) Policy for results announcement: The results were made available on the AIcrowd leaderboard and the best three results of each task were announced publicly.
Once participants submitted their results on the test set to the challenge organizers via the challenge website, they were considered fully vested in the challenge, so that their performance results (without identifying the participant unless permission was granted) became part of any presentations, publications, or subsequent analyses derived from the challenge at the discretion of the organizers.
\\(f) Publication policy: This overview paper was written by the organizing team’s members. The participating teams were encouraged to submit a paper describing their method. The participants can publish their results separately elsewhere when citing the overview paper, and (if so) no embargo will be applied. 

\paragraph{Submission method}\mbox{}\\
Submission instructions are available on the website\footnote{\url{https://www.aicrowd.com/challenges/miccai-2021-hecktor\#results-submission-format}} and are reported in the following. 
Task 1: Results should be provided as a single binary mask per patient (1 in the predicted GTVt) in .nii.gz format. The resolution of this mask should be the same as the original CT resolution and the volume cropped using the provided bounding-boxes. The participants should pay attention to saving NIfTI volumes with the correct pixel spacing and origin with respect to the original reference frame. The NIfTI files should be named [PatientID].nii.gz, matching the patient names, e.g. CHUV001.nii.gz and placed in a folder. This folder should be zipped before submission. If results are submitted without cropping and/or resampling, we will employ nearest neighbor interpolation given that the coordinate system is provided.
\\Task 2:  Results should be submitted as a CSV file containing the patient ID as "PatientID" and the output of the model (continuous) as "Prediction". An individual output should be anti-concordant with the PFS in days (i.e., the model should output a predicted risk score). 
\\Task 3: For this task, the developed methods will be evaluated on the testing set by the organizers by running them within a docker provided by the challengers. Practically, your method should process one patient at a time. It should take 3 nifty files as inputs (file 1: the PET image, file 2: the CT image, file 3: the provided ground-trugh segmentation mask, all 3 files have the same dimensions, the ground-truth mask contains only 2 values: 0 for the background, 1 for the tumor), and should output the predicted risk score produced by your model.
\\Participants were allowed five valid submissions per task. The best result was reported for each task/team. For a team submitting multiple runs to task one, the best result was determined as the highest ranking result within these runs (see ranking description in Section~\ref{subsec:task1-methods}).

\paragraph{Challenge schedule}\mbox{}
\\The schedule of the challenge, including modifications, is reported in the following.
\\\begin{itemize}
    \item the release date of the training cases: \st{June 01} June 04 2021
    \item the release date of the test cases: \st{Aug. 01} Aug. 06 2021
    \item the submission date(s): opens Sept. 01 2021 closes Sept. 10 Sept. 14 2021 (23:59 UTC-10)
    \item paper abstract submission deadline: Sept. 15 2021 (23:59 UTC-10)
    \item full paper submission deadline: Sept. 17 2021 (23:59 UTC-10)
    \item the release date of the ranking: \st{Sept. 17 2021} Sept. 27 2021
    \item associated workshop days: Sept. 27 2021
\end{itemize}

\paragraph{Ethics approval}\mbox{}
\\Montreal: CHUM, CHUS, HGJ, HMR data (training): The ethics approval was granted by the Research Ethics Committee of McGill University Health Center (Protocol Number: MM-JGH-CR15-50).
\\Lausanne: CHUV data (testing): The ethics approval was obtained from the Commission cantonale (VD) d’éthique de la recherche sur l’être humain (CER-VD) with protocol number: 2018-01513.
\\Poitiers: CHUP data (partly training and testing): The fully anonymized data originates from patients who consent to the use of their data for research purposes.

\paragraph{Data usage agreement}\mbox{}
\\The participants had to fill out and sign an end-user-agreement in order to be granted access to the data. The form can be found under the Resources tab of the HECKTOR website.

\paragraph{Code availability}\mbox{}
\\The evaluation software was made available on our github page\footnote{\url{github.com/voreille/hecktor}}.
The participating teams decided whether they wanted to disclose their code (they were encouraged to do so).

\paragraph{Conflict of interest}\mbox{}
\\No conflict of interest applies.
Fundings are specified in the acknowledgments.
Only the organizers had access to the test cases' ground truth contours.

\paragraph{Author contributions}\mbox{}

Vincent Andrearczyk: 
\\Design of the tasks and of the challenge, writing of the proposal, development of baseline algorithms, development of the AIcrowd website, writing of the overview paper, organization of the challenge event, organization of the submission and reviewing process of the participants' papers.

Valentin Oreiller: 
\\Design of the tasks and of the challenge, writing of the proposal, development of the AIcrowd website, development of the evaluation code, writing of the overview paper, organization of the challenge event, organization of the submission and reviewing process of the papers.

Sarah Boughdad:
\\Design of the tasks and of the challenge, annotations.

Catherine Cheze Le Rest:
\\Design of the tasks and of the challenge, annotations.

Hesham Elhalawani: 
\\Design of the tasks and of the challenge, annotations.

Mario Jreige: 
\\Design of the tasks and of the challenge, quality control/annotations, annotations, revision of the paper and accepted the last version of the submitted paper.

John O. Prior: 
\\Design of the tasks and of the challenge, revision of the paper and accepted the last version of the submitted paper.

Martin Vallières: 
\\Design of the tasks and of the challenge, provided the initial data and annotations for the training set~\cite{vallieres2017radiomics}, revision of the paper and accepted the last version of the submitted paper.

Dimitris Visvikis:
\\Design of the task and challenge.

Mathieu Hatt:
\\Design of the tasks and of the challenge, writing of the proposal, writing of the overview paper, organization of the challenge event.

Adrien Depeursinge: 
\\Design of the tasks and of the challenge, writing of the proposal, writing of the overview paper, organization of the challenge event.


\section{Image Acquisition Details}\label{app:acquisition}

HGJ: For the PET portion of the FDG-PET/CT scan, a median of 584 MBq (range: 368-715) was injected intravenously. After a 90-min uptake period of rest, patients were imaged with the PET/CT imaging system. Imaging acquisition of the head and neck was performed using multiple bed positions with a median of 300 s (range: 180-420) per bed position. Attenuation corrected images were reconstructed using an ordered subset expectation maximization (OSEM) iterative algorithm and a span (axial mash) of 5. The FDG-PET slice thickness resolution was 3.27 mm for all patients and the median in-plane resolution was 3.52 × 3.52 mm 2 (range: 3.52-4.69). For the CT portion of the FDG-PET/CT scan, an energy of 140 kVp with an exposure of 12 mAs was used. The CT slice thickness resolution was 3.75 mm and the median in-plane resolution was 0.98 × 0.98 mm 2 for all patients.

CHUS: For the PET portion of the FDG-PET/CT scan, a median of 325 MBq (range: 165-517) was injected intravenously. After a 90-min uptake period of rest, patients were imaged with the PET/CT imaging system. Imaging acquisition of the head and neck was performed using multiple bed positions with a median of 150 s (range: 120-151) per bed position. Attenuation corrected images were reconstructed using a LOR-RAMLA iterative algorithm. The FDG-PET slice thickness resolution was 4 mm and the median in-plane resolution was 4×4 mm 2 for all patients. For the CT portion of the FDG-PET/CT scan, a median energy of 140 kVp (range: 12-140) with a median exposure of 210 mAs (range: 43-250) was used. The median CT slice thickness resolution was 3 mm (range: 2-5) and the median in-plane resolution was 1.17 × 1.17 mm 2 (range: 0.68-1.17).

HMR: For the PET portion of the FDG-PET/CT scan, a median of 475 MBq (range: 227-859) was injected intravenously. After a 90-min uptake period of rest, patients were imaged with the PET/CT imaging system. Imaging acquisition of the head and neck was performed using multiple bed positions with a median of 360 s (range: 120-360) per bed position. Attenuation corrected images were reconstructed using an ordered subset expectation maximization (OSEM) iterative algorithm and a median span (axial mash) of 5 (range: 3-5). The FDG-PET slice thickness resolution was 3.27 mm for all patients and the median in-plane resolution was 3.52 × 3.52 mm 2 (range: 3.52-5.47). For the CT portion of the FDG-PET/CT scan, a median energy of 140 kVp (range: 120-140) with a median exposure of 11 mAs (range: 5-16) was used. The CT slice thickness resolution was 3.75 mm for all patients and the median in-plane resolution was 0.98 × 0.98 mm 2 (range: 0.98-1.37).

CHUM: For the PET portion of the FDG-PET/CT scan, a median of 315 MBq (range: 199-3182) was injected intravenously. After a 90-min uptake period of rest, patients were imaged with the PET/CT imaging system. Imaging acquisition of the head and neck was performed using multiple bed positions with a median of 300 s (range: 120-420) per bed position. Attenuation corrected images were reconstructed using an ordered subset expectation maximization (OSEM) iterative algorithm and a median span (axial mash) of 3 (range: 3-5). The median FDG-PET slice thickness resolution was 4 mm (range: 3.27-4) and the median in-plane resolution was 4 × 4 mm 2 (range: 3.52-5.47). For the CT portion of the FDG-PET/CT scan, a median energy of 120 kVp (range: 120-140) with a median exposure of 350 mAs (range: 5-350) was used. The median CT slice thickness resolution was 1.5 mm (range: 1.5-3.75) and the median in-plane resolution was 0.98 × 0.98 mm 2 (range: 0.98-1.37).

CHUV: The patients fasted at least 4h before the injection of 4 Mbq/kg of(18F)-FDG (Flucis). Blood glucose levels were checked before the injection of (18F)-FDG. If not contra-indicated, intravenous contrast agents were administered before CT scanning. After a 60-min uptake period of rest, patients were imaged with the PET/CT imaging system. First, a CT (120 kV, 80 mA, 0.8-s rotation time, slice thickness 3.75 mm) was performed from the base of the skull to the mid-thigh. PET scanning was performed immediately after acquisition of the CT. Images were acquired from the base of the skull to the mid-thigh (3 min/bed position). PET images were reconstructed by using an ordered-subset expectation maximization iterative reconstruction (OSEM) (two iterations, 28 subsets) and an iterative fully 3D (DiscoveryST). CT data were used for attenuation calculation.

CHUP: PET/CT acquisition began after 6 hours of fasting and 60±5 min after injection of 3 MBq/kg of 18F-FDG (421±98 MBq, range 220-695 MBq). Non-contrast-enhanced, non-respiratory gated (free breathing) CT images were acquired for attenuation correction (120 kVp, Care Dose® current modulation system) with an in-plane resolution of 0.853×0.853 mm2 and a 5 mm slice thickness. PET data were acquired using 2.5 minutes per bed position routine protocol and images were reconstructed using a CT-based attenuation correction and the OSEM-TrueX-TOF algorithm (with time-of-flight and spatial resolution modeling, 3 iterations and 21 subsets, 5 mm 3D Gaussian post-filtering, voxel size 4×4×4 mm3).

\newpage
%
%
%
 \bibliographystyle{splncs04}
 \bibliography{main}
\end{document}